\documentclass[sigconf,natbib=true]{acmart}
\usepackage[utf8]{inputenc}
\usepackage{amsmath}
\usepackage{amsfonts}
\usepackage{pdfpages}
\usepackage{subcaption}
\usepackage{epstopdf}
\usepackage{epsfig}
\usepackage{graphicx}
\usepackage{booktabs}
\usepackage{multirow}
\usepackage{comment}

\usepackage{algorithm}
\usepackage{algorithmic}

\usepackage{tabularx}

\usepackage{pgfgantt}
\usepackage{lscape}
\usepackage{ctable, booktabs, makecell}
\usepackage{wrapfig, layouts}
\usepackage{inputenc}
\usepackage{geometry}
\usepackage{soul}

\usepackage{color, colortbl}
\definecolor{Gray}{gray}{0.9}

\AtBeginDocument{%
  }

\setcopyright{acmlicensed}
\copyrightyear{2018}
\acmYear{2018}
\acmDOI{XXXXXXX.XXXXXXX}

\acmConference[Conference acronym 'XX]{Make sure to enter the correct
  conference title from your rights confirmation emai}{June 03--05,
  2018}{Woodstock, NY}

\acmISBN{978-1-4503-XXXX-X/18/06}

\copyrightyear{2025} 
\acmYear{2025} 
\setcopyright{cc}
\setcctype{by}
\acmConference[SIGIR '25]{Proceedings of the 48th International ACM SIGIR
Conference on Research and Development in Information Retrieval}{July 13--18,
2025}{Padua, Italy}
\acmBooktitle{Proceedings of the 48th International ACM SIGIR Conference on
Research and Development in Information Retrieval (SIGIR '25), July 13--18,
2025, Padua, Italy}\acmDOI{10.1145/3726302.3729942}
\acmISBN{979-8-4007-1592-1/2025/07}

\newcommand{\MYhref}[3][blue]{\href{#2}{\color{#1}{#3}}}

\begin{document}

\title{DARLR: Dual-Agent Offline Reinforcement Learning for Recommender Systems with Dynamic Reward}

\author{Yi Zhang}
\email{uqyzha91@uq.edu.au}
\affiliation{\institution{The University of Queensland \\ CSIRO DATA61}
  \city{Brisbane}
  \country{Australia}}
\author{Ruihong Qiu}
\email{r.qiu@uq.edu.au}
\affiliation{\institution{The University of Queensland}
  \city{Brisbane}
  \country{Australia}}
\author{Xuwei Xu}
\email{xuwei.xu@uq.edu.au}
\affiliation{\institution{The University of Queensland}
  \city{Brisbane}
  \country{Australia}}
\author{Jiajun Liu}
\email{jiajun.liu@csiro.au}
\affiliation{\institution{CSIRO DATA61 \\ The University of Queensland}
  \city{Brisbane}
  \country{Australia}}
\author{Sen Wang}
\email{sen.wang@uq.edu.au}
\affiliation{\institution{The University of Queensland}
  \city{Brisbane}
  \country{Australia}}

\renewcommand{\shortauthors}{Yi Zhang, Ruihong Qiu, Xuwei Xu, Jiajun Liu and Sen Wang}

\begin{abstract}
Model-based offline reinforcement learning (RL) has emerged as a promising approach for recommender systems, enabling effective policy learning by interacting with frozen world models. However, the reward functions in these world models, trained on sparse offline logs, often suffer from inaccuracies. Specifically, existing methods face two major limitations in addressing this challenge: (1) deterministic use of reward functions as static look-up tables, which propagates inaccuracies during policy learning, and (2) static uncertainty designs that fail to effectively capture decision risks and mitigate the impact of these inaccuracies. In this work, a dual-agent framework, DARLR, is proposed to dynamically update world models to enhance recommendation policies. To achieve this, a \textbf{\textit{selector}} is introduced to identify reference users by balancing similarity and diversity so that the \textbf{\textit{recommender}} can aggregate information from these users and iteratively refine reward estimations for dynamic reward shaping. Further, the statistical features of the selected users guide the dynamic adaptation of an uncertainty penalty to better align with evolving recommendation requirements. Extensive experiments on four benchmark datasets demonstrate the superior performance of DARLR, validating its effectiveness. The code is available at \MYhref{https://github.com/ArronDZhang/DARLR}{this address}.
\end{abstract}



\maketitle

\section{Introduction}
Recommender systems (RecSys) achieve remarkable success in industrial scenarios such as e-commerce and social media by capturing user interests to enhance their experience~\cite{bobadilla2013rs_survey,burke2002hybridRS,jiang2024challenging}. RecSys can generally be categorized into supervised learning methods~\cite{wang2024home,chang2023twin} and reinforcement learning (RL) methods. While supervised learning treats the RecSys problem as a one-off recommendation~\cite{qiu2019rethinking,qiu2020gag,li2021discovering}, making it challenging to model users' long-term dynamic interests~\cite{wanqi2023resact,wanqi2023prefrec}, RL-based methods aim to optimize long-term user satisfaction through an interactive training paradigm, a direction that has recently received increased attention~\cite{chen2023deep,chen2023opportunities}.

\begin{figure}[!t]
\centering
\begin{minipage}[b]{0.48\columnwidth}
    \includegraphics[trim=0cm 0cm 0cm 0cm, clip, width=\columnwidth]{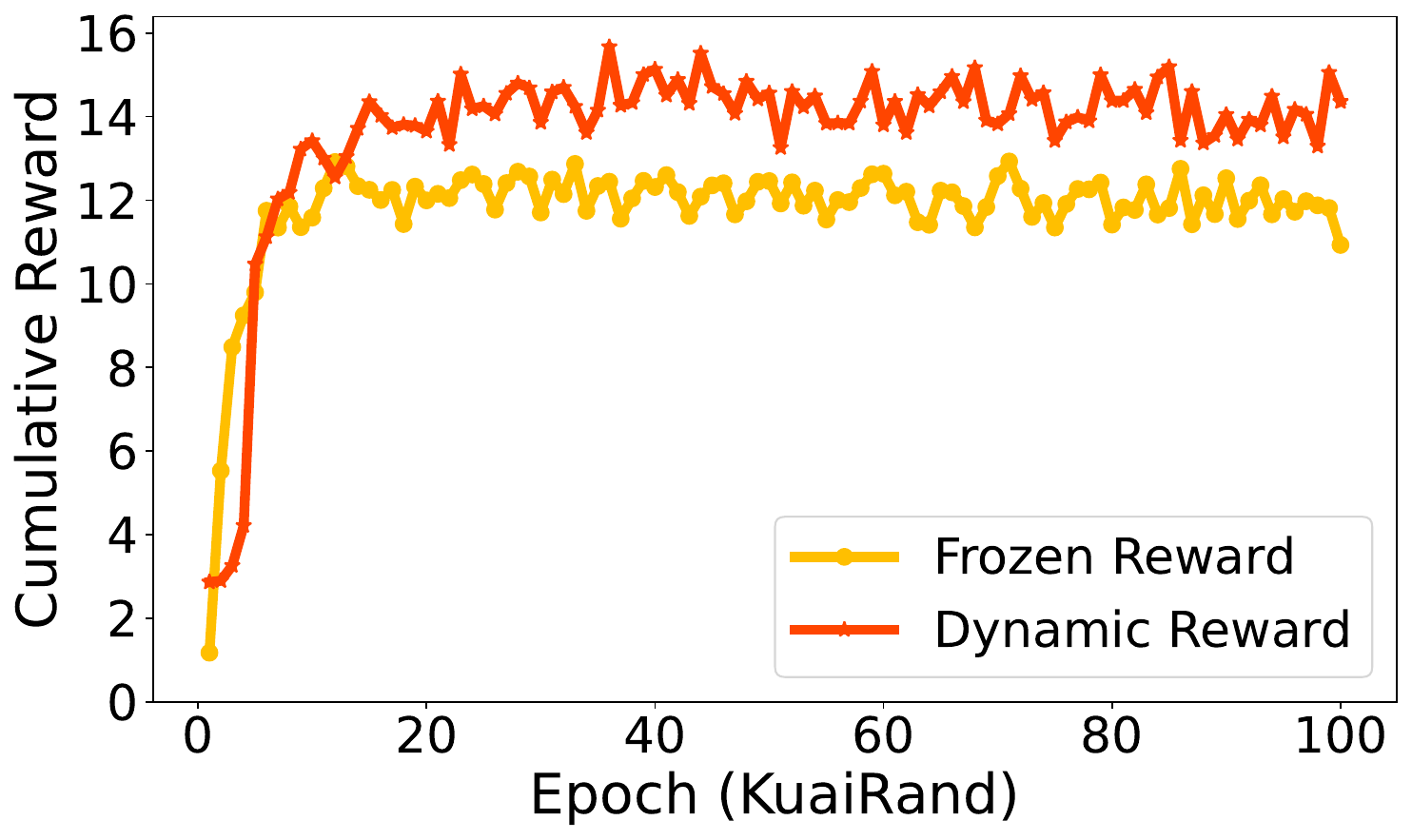}
\end{minipage}
\hfill
\begin{minipage}[b]{0.48\columnwidth}
    \includegraphics[trim=0cm 0cm 0cm 0cm, clip, width=\columnwidth]{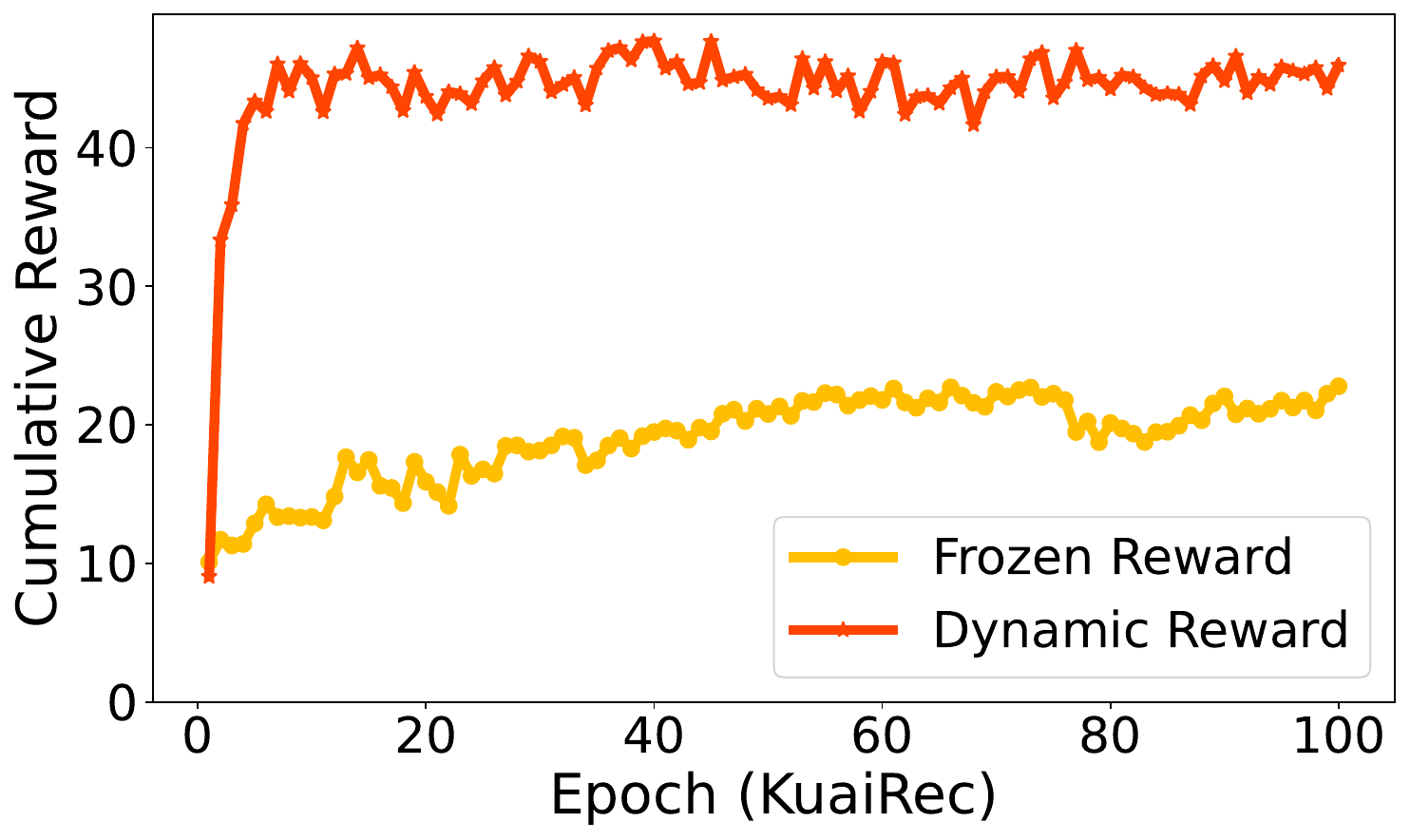}
\end{minipage}
\caption{Recommendation policies trained from scratch with dynamic reward functions converge better than those with static reward functions on KuaiRand~\cite{gao2022kuairand} and KuaiRec~\cite{gao2022kuairec}.\vspace{-0.5em}}
\label{fig:intro}
\end{figure}

Most RL-based RecSys methods~\cite{yu2024easyrl4rec,afsar2022rsrl,gao2023cirs,gao2023dorl} operate in offline RL settings, where recommendation policies are learned purely from historical user interaction logs, without requiring online interactions. Given the sparsity of these offline logs, which can hardly cover all possible user-item interactions, model-based offline RL methods~\cite{yu2021combo,janner2019mbpo,rigter2022rambo} adopt a strategy of training world models prior to policy learning. These world models include reward functions designed to predict user-item interaction rewards, typically trained using supervised learning, such as DeepFM~\cite{deepfm}, and are treated as \textit{frozen look-up tables} during recommendation policy learning~\cite{gao2023cirs,yu2024easyrl4rec}. However, these reward functions often struggle with inaccuracies, particularly for interactions that are rarely seen in the offline logs. To mitigate this, uncertainty penalties~\cite{gao2023dorl,yi2024roler} are introduced to avoid inaccurately estimated item recommendations. These penalties are also \textit{frozen} during policy learning.

Although these model-based offline RL methods for RecSys have shown promising potential in recommendation performance, their effectiveness is often hindered by: \textbf{(1) straightforward utilization} of inaccurately estimated rewards during recommendation policy learning and \textbf{(2) static uncertainty penalty} which is insufficient of capturing decision risks in RecSys.
Specifically, training recommendation policies using frozen reward functions as look-up tables inevitably amplifies the impact of inaccuracies in reward functions. For instance, overestimated items are likely prioritized during policy training~\cite{kumar2020cql,fujimoto2019bcq}, but these items often fail to meet user expectations during testing, leading to poor single-step user satisfaction and diminished long-term engagement~\cite{wanqi2023prefrec}. Meanwhile, underestimated items often tend to be overlooked, further distorting the recommendation process. Static uncertainty penalties struggle to mitigate these issues, as they fail to account for the quality of reward functions and the dynamics within the policy training process in model-based offline RL for RecSys.

Since both overestimated and underestimated rewards can harm long-term user satisfaction, iteratively refining reward functions in world models becomes essential. As shown in Fig.~\ref{fig:intro}, recommendation policies trained \textit{from scratch} with refined dynamic reward functions significantly outperform those relying on static reward functions in the long-term user satisfactory on two benchmark datasets~\cite{gao2022kuairec,gao2022kuairand}, with dynamic reward curves demonstrating faster convergence and superior performance. This underscores the feasibility and importance of dynamic reward shaping. 
To achieve this, a pioneering and effective framework, \textit{i.e.}, DARLR: \textbf{D}ual-\textbf{A}gent model-based offline \textbf{R}einforcement \textbf{L}earning for \textbf{R}eco-
mmender systems, is proposed. DARLR consists of two RL agents: the \textbf{\textit{selector}} and the \textbf{\textit{recommender}}.
The selector identifies reference users based on interactions in the world models, formulating this process as an RL task with a reward model designed to balance recommendation performance and the representativeness of selected users. The recommender then aggregates reward estimations from these reference users to refine the reward functions and adaptively estimates uncertainty penalties based on both the representativeness of the selected users and the changes in reward estimations. By \textit{continuously refining the reward estimations} of world models during the recommendation policy learning process, DARLR effectively mitigates the impact of inaccuracies in rewards. Further, as reward functions evolve, the changes of reward functions and the representativeness of selected users is dynamically updated, ensuring that the uncertainty penalty remains \textit{dynamic} and better aligned with the offline RL for RecSys setting.
The contributions are as follows:
\vspace{-1em}

\begin{itemize}
\item The limitations of inaccuracies in frozen world models within model-based offline RL for recommender systems is identified. Dynamic reward shaping is highlighted as a crucial solution to address this challenge.
\item A novel dual-agent offline RL recommender, DARLR, is proposed, which consists of a selector to identify reference users and a recommender to iteratively refine reward functions and estimate uncertainty penalties.
\item Extensive experiments are conducted on four recommendation benchmarks, the superiority of DARLR demonstrates the effectiveness of dynamic reward shaping.
\end{itemize}

\section{Preliminaries}
\noindent $\bullet$ \textbf{Reinforcement Learning} 
tasks can be formulated as Markov Decision Processes (MDPs) denoted by 5-element tuples $G = <\mathbf{S},\mathbf{A},T,r,\gamma>$~\cite{sutton2018reinforcement}, where $\mathbf{S}$ indicates the state space and each $\mathbf{s} \in \mathbf{S}$ refers to a state in the state space. $\mathbf{A}$ is the action space of the current MDP and $\mathbf{a} \in A$ represents a specific action. $T$ is the transition function, modeling the dynamics of the environment by $T(\mathbf{s_t}, \mathbf{a_t}, \mathbf{s_{t+1}}) = P(\mathbf{s_{t+1}}|\mathbf{s}=\mathbf{s_t},\mathbf{a}=\mathbf{a_t})$. $R$ is the reward function, indicating the immediate reward after taking action $\mathbf{a_t}$ at state $\mathbf{s_t}$ by $r_t = R(\mathbf{s_t}, \mathbf{a_t})$. $\gamma$ is the discount factor balancing the current immediate reward and future return. The objective of RL is to learn a policy $\pi$ that can maximize the long-term return: $G_t = \sum_{\mathbf{s}=\mathbf{s_0}}^{s_T} \gamma^t r(\mathbf{s_t}, \mathbf{a_t})$, where $\mathbf{s_0}$ stands for the initial state, $s_T$ is the last state and $t$ is the timestep. 
In addition, the state value function and state-action value function of a given policy $\pi$ are $V_{\pi}(\mathbf{s}) = \mathbf{E}_{\pi}[G_t|\mathbf{s}=\mathbf{s_t}]$ and $Q_{\pi}(\mathbf{s}, \mathbf{a}) = \mathbf{E}_{\pi}[G_t|\mathbf{s}=\mathbf{s_t}, \mathbf{a}=\mathbf{a_t}]$. 

\noindent $\bullet$ \textbf{Offline Reinforcement Learning} is a practical solution to overcome the requirement of a large number of online interactions from general RL methods~\cite{levine2020offline}. 
The offline dataset $D = \{(\mathbf{s_t}, \mathbf{a_t}, \mathbf{s_{t+1}}, r_t)\}$ is collected by one or more behavior policies denoted by $\pi_{B_i}$. 
Model-based offline RL methods usually simulate the environment using offline datasets, denoted as $G' = <\mathbf{S},\mathbf{A},\hat{T},\hat{R},\gamma>$, where $\hat{T}$ and $\hat{R}$ are the estimated transition and reward function. Then, agents interact with the learned world model to sample trajectories $\tau = \{(\mathbf{s_t}, \mathbf{a_t}, \mathbf{s_{t+1}}, \hat{r_t})'\}$. Recent approaches~\cite{yu2020mopo} incorporate uncertainty as penalties in reward functions to encourage conservatism by $r = \hat{r} - \lambda_U P_U$. This research direction exhibits substantial potential within RecSys, particularly because the inherent sparsity in offline user-item interactions significantly amplifies the risks associated with exploring extensive action spaces.

\noindent $\bullet$ \textbf{Formulating Recommendation as a Reinforcement Learning Problem} is often referred to as RL4RS.
\label{rl4rs_form}
Recalling the 5-element tuple of MDP, each state \(\mathbf{s} \in \mathbf{S}\) corresponds to a status that the RecSys needs to recommend a specific item to a user. A $\mathbf{s}$ usually comprises the user's side information, such as gender and age, and recent interaction history~\cite{huang2022state_repr}. An action $\mathbf{a}$ corresponds to one item in the item set denoted by $\mathbf{A}$. The reward \(r(s_t, a_t)\) is derived from the feedback after recommending item \(\mathbf{a_t}\) in state \(\mathbf{s_t}\), and it varies depending on the dataset, such as watch time on a short video platform or ratings on a product review site. As for the transition function, it serves as a state tracker, capturing the sequence of states autoregressively: \(\mathbf{s'} = f(\mathbf{s},\mathbf{a},r)\), and is often implemented using sequential models~\cite{kang2018sasrec,wei2024leave,zhou2024contrastive}. Ultimately, the objective of the RecSys is to learn a policy \(\pi\) that maximizes the long-term user experience, expressed as \(\arg \max_{\pi} \mathbf{E}_{\tau \sim \pi}[\sum_{(\mathbf{s},\mathbf{a}) \in \tau} \gamma^t r(\mathbf{s}, \mathbf{a})]\), where \(\tau\) are trajectories generated by policy \(\pi\). This paper focuses on model-based offline RL, adhering to state-of-the-art approaches, such as DORL~\cite{gao2023dorl} and ROLeR~\cite{yi2024roler}, to comprehensively describe the learning process. These approaches typically involve two phases: learning the world model from offline interaction, followed by training the recommendation policy within this simulated environment.

\section{Method}
\begin{figure}[!t]
\centering
\includegraphics[trim=0cm 0cm 0cm 0cm, clip, width=1\linewidth]{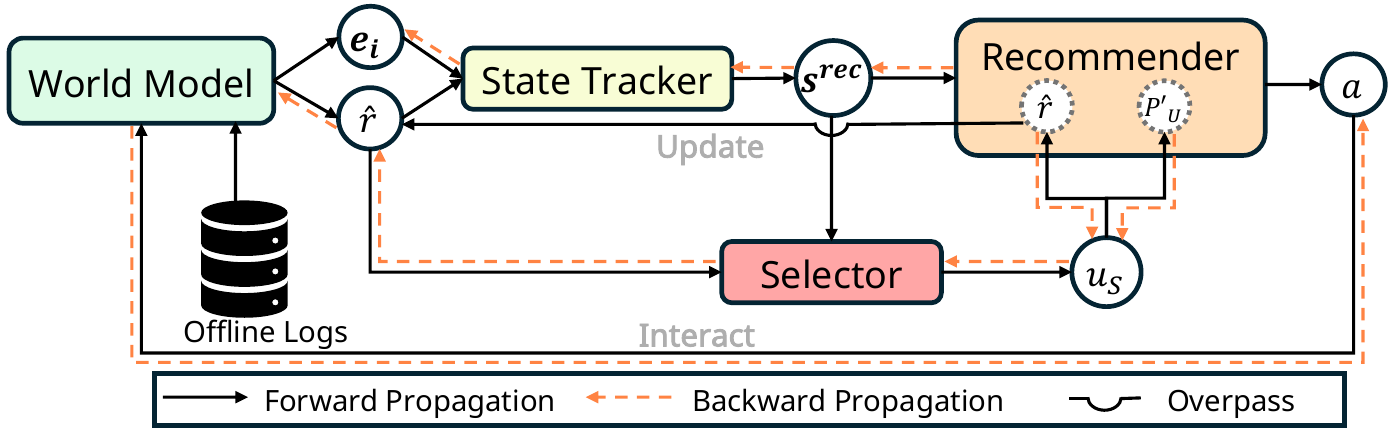}
\vspace{-2em}
\caption{
Overall Framework of DARLR with three main modules: a world model, a selector agent and a recommender agent. The gradient flow will influence the training of both agents, as well as the update of the world model.}
\label{fig:framework}
\vspace{-1.5em}
\end{figure}

Generally, DARLR comprises three main modules: the world model, the selector, and the recommender, as shown in Figure~\ref{fig:framework}. 

\subsection{World Model Learning}
Intuitively, a world model is a simulated recommendation system. It is learned with offline data through supervised learning. Specifically, during world model learning, users and items are encoded from IDs and features into embeddings. The reward function in the world model is trained through a predictive model such as DeepFM~\cite{deepfm}. 
In addition, some types of uncertainty penalty~\cite{gao2023cirs,yu2024easyrl4rec} and entropy penalty~\cite{gao2023dorl} are also calculated in this stage.

\textbf{Item Embedding}, $\mathbf{e}_{i} \in \mathbb{R}^{d_I}$, is derived from the item ID and part of item features such as tags for music or categories for products. 
\begin{equation}
    \abovedisplayskip=2pt
    \belowdisplayskip=2pt
    \mathbf{e}_{i} = f_\text{Item}(i_{id}, \mathbf{F}^i_{id}), 
\end{equation}
where $f_\text{Item}$ represents the item encoder, and $i_{id}$, $\mathbf{F^i}$ denotes the item id and item features, respectively.

\textbf{User Embedding}, $\mathbf{e}_u\in \mathbb{R}^{d_U}$, analogous to item embedding, utilizing user ID and auxiliary information.
\begin{equation}
    \abovedisplayskip=2pt
    \belowdisplayskip=2pt
    \label{item_emb}
    \mathbf{e}_u = f_\text{User}(u_{id}, \mathbf{F}^u_{id}),
\end{equation}
where $f_\text{User}$ is the user encoder, and $u_{id}$, $\mathbf{F}^u$ represents the user id and the user features, respectively. $\mathbf{e}_u\in \mathbb{R}^{d_U}$ is time-invariant here.

\textbf{Reward Prediction} is a key output of the world model. In RecSys, user-item interaction logs can be regarded as a sparse matrix with each row corresponding to one user and each column corresponding to one item. The $(i, j)-$th element in the matrix represents user $i$'s feedback about item $j$. After supervised learning, the sparse matrix is filled with predicted feedback predictions, \textit{i.e.,} $\hat{r} \in \mathbb{R}$. Though these predictions complete the reward function, especially for the unseen user-item interactions in offline data, the influence of the accuracy of the reward predictions on the recommendation policy learning still needs exploring. DORL~\cite{gao2023dorl} and CIRS~\cite{gao2023cirs} average predictions from multiple world models (WMs) to estimate rewards by:
\begin{equation}
    \abovedisplayskip=2pt
    \belowdisplayskip=2pt
    \label{r1}
    \hat{r} = \frac{1}{K} \sum^K_{j=1} \text{WM}_j(\mathbf{e}_i, \mathbf{e}_u),
\end{equation} 
where $K$ is the number of world models, and $\text{WM}_j$ represents the $j$-th world model.

\textbf{Uncertainty Penalty} is a crucial component in offline RL, often utilized to evaluate the risk associated with specific actions. Some approaches~\cite{yu2020mopo,kidambi2020morel} utilize an ensemble of world models to estimated the uncertainty. In the prior work, DORL~\cite{gao2023dorl}, it uses Gaussian Probabilistic Model (GPM) to facilitate the building of world models, with the uncertainty of an interaction $x$ calculated as $P_U(x) := \max_{k \in E} \sigma^2_{\theta_k}$, where \(k\) indexes the world model ensemble $E$ and \(\sigma^2_{\theta_k}\) represents the variance of the corresponding GPM. Consequently, the reward estimate in Equation (\ref{r1}) is modified as:
\begin{equation}
    \abovedisplayskip=2pt
    \belowdisplayskip=2pt
    \label{r2}
    \bar{r}' = \hat{r} - \lambda_U P_U,
\end{equation}
where \(\lambda_U\) is the uncertainty coefficient that scales the penalty. Based on the formulation of the uncertainty penalty, the quality of world models has significant influence on the estimation of uncertainty. To release the impact, a new uncertainty penalty tailored for offline RL4RS is proposed in Sec.~\ref{sec:uncertain}.

\textbf{Entropy Penalty} is one of the key contributions of DORL to alleviate the Matthew effect. It is derived from the offline data, independent of the world modeling learning. The entropy penalty is calculated as the sum of a \(k\)-order entropy, which considers the recent \(k\) interactions as a pattern and evaluates the likelihood of the next item aligning with the current pattern:
\begin{equation}
    \abovedisplayskip=2pt
    \belowdisplayskip=2pt
    \label{pe}
    P_E = -D_{\text{KL}}(\pi_{\beta}(\cdot|\mathbf{s})||\pi_u(\cdot|\mathbf{s})),
\end{equation}
where \(\pi_{\beta}\) denotes the behavior policy and \(\pi_u\) the uniform distribution. This penalty effectively encourages policy exploration, enhancing cumulative rewards by getting rid of suboptimal. The final reward function of DORL for policy learning is:
\begin{equation}
    \abovedisplayskip=2pt
    \belowdisplayskip=2pt
    r = \hat{r} - \lambda_U P_U + \lambda_E P_E,
\end{equation}
where \(\lambda_E\) is the coefficient for entropy. It controls the scale of exploration is remained in this study due to its effectiveness.

\subsection{Selector}
\label{selector}
The selector searches for reference uses for a given user by leveraging the intuition that users' preferences can be \textbf{dynamically represented} by that of similar users.

\subsubsection{Selector Modeling} The RL formulation is detailed here.
\label{sec:selector}

\textbf{State Representation}. During recommendation policy learning, users interact with the RecSys. Within each step of user-item interaction, there are multiple selection steps. Thus, the state for the selector needs to consider information of two parts: recommendation steps and selection steps. Specifically, features from recommendation steps consist of the user embedding in Equation (\ref{item_emb}) and recent interacted items. It is captured by the state tracker of the recommender which will be introduced in Sec.~\ref{sec:rec_modeling} and it is denoted as $\mathbf{s^{\text{rec}}_t} \in \mathbb{R}^{d_R}$. Meanwhile, features from the selection steps derive from selected users' preferences. The user-item feedbacks predicted from the world models are used here, represented as $\mathbf{p_u} \in \mathbb{R}^{d}$, \textit{i.e.,} $\mathbf{p_u}$ corresponds to one row of the predicted user-item feedback matrix. So the initial selector state for user $u$ is given by:
\begin{equation}
    \abovedisplayskip=2pt
    \belowdisplayskip=0pt
    \mathbf{s^{\text{sel}}_{u,0}} = \mathbf{s^{\text{rec}}_t} \oplus \mathbf{p_u},
\end{equation}
where $\oplus$ indicates concatenation and $\mathbf{s^{\text{sel}}_{u,0}} \in \mathbb{R}^{d_S}$. As for the transition function of the selector, it progressively incorporates the newly added users' feedbacks with a sequence model~\cite{vaswani2017attention}:
\begin{equation}
    \abovedisplayskip=2pt
    \belowdisplayskip=2pt
    \label{sel_state}
    \mathbf{s^{\text{sel}}_{u,t+1}} = \text{Transformer}(\mathbf{s^{\text{sel}}_{u,t-w^{sel}+1}}, \cdots, \mathbf{s^{\text{sel}}_{u,t}}),
\end{equation}
where $w^{sel}$ denotes the window size of the selector's transition function. The selection terminates when $K^{sel}$ users' preferences are collected. It is a hyperparameter related to the size and sparsity of the dataset, which will be discussed in the experiment section.

\textbf{User Selection.} In each selection step, one user who is supposed to be a reference of the current user is selected:
\begin{equation}
    \abovedisplayskip=2pt
    \belowdisplayskip=2pt
    \label{eq:select}
    u_t = \pi_{sel}(\mathbf{s^{\text{sel}}_{u,t}}),
\end{equation}
where $\pi_{sel}$ refers to the selection policy.

\textbf{Dynamic Selector Reward Design.} In an online reinforcement learning setting, the reward design for the selector can be intuitive: it shares the same recommendation reward function from the RecSys~\cite{luo2019curiosity}. Unfortunately, the reward design poses a substantial challenge in an offline setting since it is difficult to learn selection policy by interacting with the simulated world models. In other words, the simulated world model cannot directly \textit{teach} the selector whether the selected users are representative enough of the current users. Therefore, an intrinsic reward function designed for the selector is urgent. Three aspects are considered in the proposed selection reward model at selection step $t$ for current user $u$: as maximizing the cumulative recommendation reward is the ultimate target, single-step recommendation reward, \textit{i.e.,} $\hat{r}$ (initially obtained from Equation (\ref{r1})), is part of the intrinsic reward. Besides, based on that similar users may share similar preferences~\cite{yi2024roler,wang2020aspect}, the similarities between the selected user' feedback vector, \textit{i.e.,} $p_u$, and that of the current user are another part, termed as similarity gain:
\begin{equation}
    \abovedisplayskip=2pt
    \belowdisplayskip=2pt
    \label{r_s^sel}
    r^{sel}_{s} = \text{cos}(\mathbf{p_u}, \mathbf{p_{u_t}}),
\end{equation}
where $\text{cos}$ denotes cosine similarity and $r_s^{\text{sel}} \in \mathbb{R}$.
Additionally, to ensure that the selected users are representative and informative, the selected users are expected to be diverse. Thus, the selected user's dissimilarities with the already selected users are the third part of the reward, termed as diversity gain:
\begin{equation}
    \abovedisplayskip=2pt
    \belowdisplayskip=2pt
    \label{r_d^sel}
    r^{sel}_{d} = \frac{\sum_{u_i \in u_S} [1-\text{cos}(u_i, u_t)]}{||u_S||},
\end{equation}
where $u_S$ is the selected user set, $||u_S||$ is the cardinal number of this set and $r_d^{\text{sel}} \in \mathbb{R}$. When $||u_S|| = 0$, $r^{sel}_{d}=0$. Combining the three parts, the intrinsic reward model is:
\begin{equation}
    \abovedisplayskip=2pt
    \belowdisplayskip=2pt
    \label{eq:intrinsic_r}
    r^{sel} = \hat{r} + \lambda_s r^{sel}_{s} + \lambda_d r^{sel}_{d},
\end{equation}
where $\lambda_s$ and $\lambda_d$ are two hyperparameters controlling the scale of the corresponding parts.

\subsubsection{Selector Training}
It is believed that the proposed selector is general so that it can be implemented by a wide range of policy gradient and actor-critic RL algorithms. For simplicity, Advantage actor-critic (A2C)~\cite{mnih2016a2c} is chosen in DARLR. The selection happens within each recommendation step. Given a user $u$, the actor sequentially selects reference users for $u$ aiming to maximize the value function of the critic until the termination of selection during the forward process. Then, the critic's value estimation is updated during loss back-propagation. It is noteworthy that employing static reward shaping, such as ROLeR~\cite{yi2024roler}, as an initialization step can effectively mitigate the risk of propagating uncorrected early-stage selector errors throughout subsequent dynamic reward iterations.

\subsection{Recommender}
\label{pipline}
The recommendation agent manages interactions between users and the RecSys. It learns a recommendation policy to enhance long-term user engagement and experience. The general formulation is described in Sec.~\ref{rl4rs_form}. In this part, the detailed modeling of the recommender in DARLR and the connection between the selector and recommender are exhibited.

\subsubsection{Recommender Modeling} 
\label{sec:rec_modeling}
For readability, the modeling introduction begins with action representation, followed by state representation and the transition function. Subsequently, the uncertainty penalty design and dynamic reward modeling are presented.

\textbf{Action Representation.}
For each recommendation step, the current user is recommended with an item. Intuitively, the item embedding is used as the action representation~\cite{gao2023cirs,gao2023dorl,yi2024roler}:
\begin{equation}
    \abovedisplayskip=2pt
    \belowdisplayskip=2pt
    \label{act_repr}
    \mathbf{a_t} =  \mathbf{e}_{i}.
\end{equation}

\textbf{State Tracker} captures the dynamics of the RecSys, \textit{i.e.,} modeling the next state representation after a user interacts with an item in the current state. Thus, it also serves as the state encoder for the recommender, which can be formulated as $\mathbf{s^{rec}_{t+1}} = P^{rec}(\mathbf{s^{rec}_{t}}, \mathbf{a_t})$. Following the design of DORL~\cite{gao2023dorl}, the state tracker takes recent interacted items and received rewards into account. Further, owing to the sequential interaction nature, Transformer~\cite{vaswani2017attention} tracker~\cite{kang2018sasrec,yi2024roler} is preferred to seize the ordering relationship instead of the average tracker in DORL:
\begin{equation}
    \abovedisplayskip=2pt
    \belowdisplayskip=2pt
    \label{rec_state}
    \mathbf{s^{\text{rec}}_{t+1}} = \text{Transformer}(\mathbf{s^{\text{rec}}_{t-w^{rec}+1}}, \cdots, \mathbf{s^{\text{rec}}_{t}}),
\end{equation}
where $w^{rec}$ is the window size of the recommender's transition function. Note that the state modeling of the recommender is similar to that of the selector, \textit{i.e.,} Equation (\ref{sel_state}), except that the selector's state encoder captures information of two granularities initially.

\textbf{Dynamic Recommender Reward Modeling.} 
To decrease the impact of the world model upon recommendation policy learning, an effective offline reward shaping method is necessary. Though ROLeR~\cite{yi2024roler} proposes an non-parametric clustering based method, it is a one-time reward shaping and is static during training, which does not fully utilize the offline data and discover the potential relationships between users' preferences. Therefore, the selector depicted in Sec.~\ref{selector} is in charge of finding reference users of the current user at each recommendation step to continuously improve the reward models of the world models. Since the process description will not go into specific selection steps, $t$ refers to the recommendation step in this part. After selection, all reference users gathers a user-dependent and step-dependent set $u_{S,t}$. Then, the dynamic reward shaping can be implemented by averaging over the set:
\begin{equation}
    \abovedisplayskip=2pt
    \belowdisplayskip=2pt
    \label{reward_shaping}
    \hat{r}(u, i_t) = \frac{\sum_{u' \in u_{S,t}}\hat{r}(u',i_t)}{||u_{S,t}||},
\end{equation}
where $i_t$ refers to the item selected at step $t$. Note that the same notation as Equation (\ref{r1}) is used here. It implies that the reward shaping is conducted along the whole learning process to constantly easing the impact of the inaccuracy within world models.

\vspace{-1em}
\subsubsection{Uncertainty Design} \label{sec:uncertain} Uncertainty penalty is a widely accepted technique used in offline RL~\cite{levine2020offline,chen2023opportunities} to mitigate the possibilities of choosing risky actions and encouraging conservative policies. Though diverse uncertainty penalties are proposed, most of them are not tailored for RecSys. The extra sparse offline data in RecSys poses special difficulties in controlling the degree of conservatism. To address this, DARLR designs a simple yet effective uncertainty penalty that flexibly estimates the risk of the current action. This uncertainty penalty exploits the similarity gain, \textit{i.e.,} $r^{sel}_s$ from Equation (\ref{r_s^sel}), and diversity gain, \textit{i.e.,} $r^{sel}_d$ from Equation (\ref{r_d^sel}):
\begin{equation}
    \abovedisplayskip=0pt
    \belowdisplayskip=0pt
    \label{uncertainty}
    P_U' = \frac{|\hat{r}-\hat{r}_{-1}|}{r^{sel}_s + r^{sel}_d},
\end{equation}
where $\hat{r}_{-1}$ refers to the last reward estimation. The motivation of this design can be expressed by the representativeness of the reference user set, \textit{$u_S$}.
When reward predictions deviate significantly from the previous iteration, a sharp change is acceptable if the similarity and diversity gains indicate that the selected set is sufficiently representative. However, if these two values fail to ensure representativeness, such abrupt changes introduce risks, and the selection probabilities for the corresponding actions should be constrained accordingly. The uncertainty penalty can also dynamically evolve during policy learning to ensure adaptability instead of being static in existing work~\cite{gao2023dorl,yi2024roler,gao2023cirs}.

To sum up, the ultimate reward function for training the recommendation policy is described as:
\begin{equation}
    \abovedisplayskip=2pt
    \belowdisplayskip=2pt
    \label{eq:rec_reward}
    r^{rec} = \hat{r} - \lambda_U P_U' + \lambda_E P_E.
\end{equation}
The entropy penalty from Equation (\ref{pe}) is remained to encourage recommendation diversity.

\subsubsection{Recommender Training} As choosing a specific RL algorithm is not the focus of this work, and to keep consistency and fair comparisons, A2C is adopted for training the policies~\cite{gao2023dorl,gao2023cirs,yi2024roler}.

\begin{algorithm}[!t]
\caption{Dual-Agent offline RL Recommender (DARLR).}
\label{alg:Framwork}
\begin{algorithmic}[1] 
\REQUIRE ~~\\ 
    Learned world model $E$ (Environment); Training epoch of the recommender $K$; Number of trajectories in each epoch $N$; 
\ENSURE ~~\\ 
   Recommendation policy, $\pi^{R}_\theta$;
    \STATE Initialize the actor and critic network parameters of the recommender and the selector, $\theta^R$, $\phi^R$, $\theta^S$ and $\phi^S$;
    \FOR{each epoch $k = 0,1,2, ..., K$}
    \STATE n=0
    \WHILE{n < N}
        \FOR{each recommendation step}
        \STATE The actor of the selector samples a trajectory $\tau^S$ with current policy $\pi_{\theta}^S$ and stores it in $T$;
        \ENDFOR
    \STATE The actor samples a trajectory $\tau^R$ with current policy $\pi_{\theta}^R$ by interacting with the environment $E$;
    \STATE Calculate advantages of the selector and the recommender with $A(\mathbf{s_t},\mathbf{a_t})=Q(\mathbf{s_t},\mathbf{a_t})-V(\mathbf{s_t})$ and the cumulative reward for each selection step in $T$ and each recommendation step in $\tau^{\text{sel}}$, respectively;
    \STATE Update the critics for the selector and the recommender \textit{i.e.,} $\phi^S$ and $\phi^R$, with loss calculated as $L_{\text {critic }}=\mathbb{E}_\tau\left[\left(r_t+\gamma \max _{\mathbf{a}^{\prime}} Q\left(\mathbf{s}_{\mathbf{t}+\mathbf{1}}, \mathbf{a}^{\prime} ; \phi\right)-Q\left(\mathbf{s}_{\mathbf{t}}, \mathbf{a}_{\mathbf{t}} ; \phi\right)\right)^2\right]$;
    \STATE Update the actor, \textit{i.e.,} $\theta^R$ and $\theta^S$, by ascending the policy gradient of $J_{\text {actor }}=\mathbb{E}_\tau\left[\log \pi_\theta\left(\mathbf{a}_{\mathbf{t}} \mid \mathbf{s}_{\mathbf{t}}\right) \cdot A\left(\mathbf{s}_{\mathbf{t}}, \mathbf{a}_{\mathbf{t}}\right)\right]$;
    \STATE $n = n+1$;
    \ENDWHILE
    \ENDFOR
\RETURN the policy $\pi_\theta^R$; 
\end{algorithmic}
\end{algorithm}

\subsection{Learning Framework of DARLR}
To recap and enhance the readability, the overall learning of DARLR is illustrated in Alg.~\ref{alg:Framwork}. Note that the superscript of advantage functions, actors and critics are omitted for brevity. During the forward process, DARLR's dual-agent begins sampling. Given a user, the actor of the recommender recommends an item with the aim to maximize its critic. Then, the selector's actor progressively selects reference users for the current user guided by its critic. The corresponding reward prediction and uncertainty penalty are obtained after selection. In the backward process, temporal-difference (TD) losses are calculated for the critics of the recommender and the selector. And advantage gradient losses are calculated for both actors. After the losses being back-propagated, the selector and recommender are updated.

\section{Experiments}
In this section, experiments will be conducted to evaluate the effectiveness of DARLR with the following research questions (RQs):

\noindent$\bullet$ (RQ1) How does DARLR perform compared with other baselines?

\noindent$\bullet$ (RQ2) How does the selector, including its design and components, contribute to DARLR's performance?

\noindent$\bullet$ (RQ3) What are the benefits of the dynamic reward functions?

\noindent$\bullet$ (RQ4) Is DARLR robust to the critical hyperparameters? 

\begin{table}[!t]
    \centering
    \caption{Dataset statistics.}
    \vspace{-1em}
    \label{tb:dataset}
    {
    \renewcommand{\arraystretch}{0.3}
    \begin{tabular}{ccccc}
    \toprule
    Dataset                   & Usage & \# Users & \# Items & \makecell{Density}   \\
    \toprule
    \multirow{2}{*}{KuaiRand} & Train & 27285        & 7551         & 0.697\%              \\
                              & Test  & 27285        & 7583         & 0.573\%             \\
    \midrule
    \multirow{2}{*}{KuaiRec}  & Train & 7176         & 10728        & 16.277\%             \\
                              & Test  & 1411         & 3327         & 99.620\%             \\
    \midrule
    \multirow{2}{*}{Coat}     & Train & 290          & 300
       & 8.046\%              \\
                              & Test & 290           & 300
       & 5.287\%              \\
    \midrule
    \multirow{2}{*}{Yahoo}  & Train & 15400        & 1000
       & 2.024\%              \\
                              & Test & 5400          & 1000
       & 1.000\%              \\
    \bottomrule
    \end{tabular}
    }
    \vspace{-1em}
\end{table}

\subsection{Setup}
\subsubsection{Datasets}
Experiments are conducted on four benchmark datasets following~\cite{gao2023dorl,gao2023cirs,yi2024roler} with statistics in Table~\ref{tb:dataset}.

\noindent $\bullet$ \textbf{KuaiRand}~\cite{gao2022kuairand} contains short video browsing histories which is very sparse and collected by randomly exposing target videos to users in the standard recommendation streams.

\noindent $\bullet$ \textbf{KuaiRec}~\cite{gao2022kuairec} contains short video platform browsing histories with a fully observable user-item interaction matrix where users' feedback on items is known.

\noindent $\bullet$ \textbf{Coat}~\cite{schnabel2016coat} consists of shopping ratings where products can be categorized into user-self-selected and uniformly sampled items.

\noindent $\bullet$ \textbf{Yahoo}~\cite{marlin2009yahoo} consists of music rating records with user-selected songs for training and records of randomly picked songs for testing.

\begin{table*}[!t]
\caption{The performance summary of all methods on KuaiRand and KuaiRec. GT Reward serves as the upper bound using the ground-truth (GT) reward to train the RL agent. MCD is a reference that should be controlled instead of the lower the better~\cite{gao2023dorl}. (Bold: best; Underline: runner-up)}
\vspace{-0.7em}
\label{tb:main_result1}
\centering
\resizebox{\linewidth}{!}{
\begin{tabular}{l|c|c|c|c|c|c|c|c}
\toprule
\multirow{2}{*}{Method} & \multicolumn{4}{c|}{KuaiRand} & \multicolumn{4}{c}{KuaiRec} \\
\cmidrule{2-5}
\cmidrule{6-9}
 & $R_\text{tra}$ $\uparrow$ & $R_\text{each}$ $\uparrow$ & Length $\uparrow$ & MCD & $R_\text{tra}$ $\uparrow$ & $R_\text{each}$ $\uparrow$ & Length $\uparrow$ & MCD \\
\midrule
\midrule
UCB & $1.6510\pm 0.1515$ & $0.3725\pm 0.0278$ & $4.4312\pm 0.2121$ & $0.7886\pm 0.0235$ & $3.6059\pm 0.6092$ & $0.8531\pm 0.1145$ & $4.2190\pm 0.3892$ & $0.8112\pm 0.0582$ \\
$\epsilon$-greedy & $1.7109\pm 0.1258$ & $0.3510\pm 0.0251$ & $4.8804\pm 0.2700$ & $0.7735\pm 0.0239$ & $3.5152\pm 0.7315$ & $0.8276\pm 0.1286$ & $4.2186\pm 0.4049$ & $0.8226\pm 0.0482$ \\
SQN & $0.9117\pm 0.9292$ & $0.1818\pm 0.0584$ & $4.6007\pm 3.7125$ & $0.6208\pm 0.1865$ & $4.6730\pm 1.2149$ & $0.9125\pm 0.0551$ & $5.1109\pm 1.2881$ & $0.6860\pm 0.0931$ \\
CRR & $1.4812\pm 0.1236$ & $0.2258\pm 0.0151$ & $6.5613\pm 0.3519$ & $0.7326\pm 0.0187$ & $4.1631\pm 0.2535$ & $0.8945\pm 0.0365$ & $4.6541\pm 0.2150$ & $0.8648\pm 0.0168$ \\
CQL & $2.0323\pm 0.1070$ & $0.2258\pm 0.0119$ & $9.0000\pm 0.0000$ & $0.7778\pm 0.0000$ & $2.5062\pm 1.7665$ & $0.6843\pm 0.2279$ & $3.2239\pm 1.3647$ & $0.3858\pm 0.3853$ \\
BCQ & $0.8515\pm 0.0523$ & $0.4246\pm 0.0164$ & $2.0050\pm 0.0707$ & $0.9983\pm 0.0236$ & $2.1234\pm 0.0815$ & $0.7078\pm 0.0272$ & $3.0000\pm 0.0000$ & $0.6667\pm 0.0000$ \\
MBPO & $10.9325\pm 0.9457$ & $0.4307\pm 0.0210$ & $25.3446\pm 1.8190$ & $0.3061\pm 0.0403$ & $12.0426\pm 1.3115$ & $0.7701\pm 0.0290$ & $15.6461\pm 1.6373$ & \underline{$0.3621\pm 0.0465$} \\
IPS & $3.6287\pm 0.6763$ & $0.2163\pm 0.0141$ & $16.8213\pm 3.1824$ & $\mathbf{0.2010\pm 0.1156}$ & $12.8326\pm 1.3531$ & $0.7673\pm 0.0234$ & $16.7270\pm 1.6834$ & $\mathbf{0.2150\pm 0.0644}$ \\
MOPO & $10.9344\pm 0.9634$ & $0.4367\pm 0.0193$ & $25.0019\pm 1.8911$ & $0.3433\pm 0.0289$ & $11.4269\pm 1.7500$ & $0.8917\pm 0.0505$ & $12.8086\pm 1.8502$ & $0.4793\pm 0.0619$ \\
DORL & $11.8500\pm 1.0361$ & $0.4284\pm 0.0223$ & $27.6091\pm 2.1208$ & \underline{$0.2960\pm 0.0356$} & $20.4942\pm 2.6707$ & $0.7673\pm 0.0264$ & $26.7117\pm 3.4190$ & $0.3792\pm 0.0149$ \\
ROLeR & \underline{$13.4553 \pm 1.5086$} & \underline{$0.4574\pm 0.0332$} & $\mathbf{29.2700\pm 2.3225}$ & $0.4049\pm 0.0356$ & \underline{$33.2457\pm 2.6403$} & \underline{$1.2293\pm 0.0511$} & \underline{$27.0131\pm 1.3986$} & $0.4439\pm 0.0212$ \\
\midrule
\rowcolor{lightgray}
DARLR (Ours) & $\mathbf{13.8152 \pm 1.9351}$ & $\mathbf{0.4670 \pm 0.0445}$ & \underline{$29.2028 \pm 2.3869$} & $0.4178 \pm 0.0411$ & $\mathbf{35.2203 \pm 2.7576}$ & $\mathbf{1.2600 \pm 0.0404}$ & $\mathbf{27.3526 \pm 2.1905}$ & $0.4869 \pm 0.0438$ \\
\midrule
\midrule
\textcolor{gray}{GT (Ideal)} & \textcolor{gray}{$14.3689\pm 1.9708$} & \textcolor{gray}{$0.4993\pm 0.0488$} & \textcolor{gray}{$28.5582\pm 2.4114$} & \textcolor{gray}{$0.4109\pm 0.0397$} & \textcolor{gray}{$36.7475\pm 3.4738$} & \textcolor{gray}{$1.5600\pm 0.0405$} & \textcolor{gray}{$23.5653\pm 2.1824$} & \textcolor{gray}{$0.5594\pm 0.0267$} \\
\bottomrule
\end{tabular}
}
\end{table*}

\begin{table*}[!t]
\caption{The performance summary of all methods on Coat and Yahoo. GT Reward serves as the upper bound using the ground-truth (GT) reward to train the RL agent. (Bold: best; Underline: runner-up). The MCD metric is not applicable for Coat and Yahoo since both datasets do not include the attribute of \textit{most popular categories}.}
\vspace{-0.7em}
\label{tb:main_result2}
\centering
\resizebox{0.8\linewidth}{!}{
\begin{tabular}{l|c|c|c|c|c|c}
\toprule
\multirow{2}{*}{Method} & \multicolumn{3}{c|}{Coat} & \multicolumn{3}{c}{Yahoo} \\
\cmidrule{2-4}
\cmidrule{5-7}
 & $R_\text{tra}$ $\uparrow$ & $R_\text{each}$ $\uparrow$ & Length $\uparrow$ & $R_\text{tra}$ $\uparrow$ & $R_\text{each}$ $\uparrow$ & Length $\uparrow$ \\
\midrule
\midrule
UCB & $73.6713\pm 1.8105$ & $2.4557\pm 0.0604$ & $30.0000\pm 0.0000$ & $66.7578\pm 1.2539$ & $2.2253\pm 0.0418$ & $30.0000\pm 0.0000$ \\
$\epsilon$-greedy & $72.0042\pm 1.6054$ & $2.4001\pm 0.0535$ & $30.0000\pm 0.0000$ & $64.3439\pm 1.2911$ & $2.1448\pm 0.0430$ & $30.0000\pm 0.0000$ \\
SQN & $72.6142\pm 2.0690$ & $2.4205\pm 0.0690$ & $30.0000\pm 0.0000$ & $57.7270\pm 5.7506$ & $1.9242\pm 0.1917$ & $30.0000\pm 0.0000$ \\
CRR & $67.3830\pm 1.6274$ & $2.2461\pm 0.0542$ & $30.0000\pm 0.0000$ & $57.9941\pm 1.6752$ & $1.9331\pm 0.0558$ & $30.0000\pm 0.0000$ \\
CQL & $68.9835\pm 1.8659$ & $2.2995\pm 0.0622$ & $30.0000\pm 0.0000$ & $62.2909\pm 3.3466$ & $2.0764\pm 0.1116$ & $30.0000\pm 0.0000$ \\
BCQ & $68.8012\pm 1.7627$ & $2.2934\pm 0.0588$ & $30.0000\pm 0.0000$ & $61.7388\pm 1.7808$ & $2.0580\pm 0.0594$ & $30.0000\pm 0.0000$ \\
MBPO & $71.1930\pm 2.0943$ & $2.3731\pm 0.0698$ & $30.0000\pm 0.0000$ & $64.5500\pm 2.1567$ & $2.1517\pm 0.0719$ & $30.0000\pm 0.0000$ \\
IPS & $73.8872\pm 1.8417$ & $2.4629\pm 0.0614$ & $30.0000\pm 0.0000$ & $57.8499\pm 1.7955$ & $1.9283\pm 0.0599$ & $30.0000\pm 0.0000$ \\
MOPO & $71.1805\pm 2.0560$ & $2.3727\pm 0.0685$ & $30.0000\pm 0.0000$ & $65.5098\pm 2.0996$ & $2.1837\pm 0.0700$ & $30.0000\pm 0.0000$ \\
DORL & $71.3992\pm 2.0640$ & $2.3800\pm 0.0688$ & $30.0000\pm 0.0000$ & $66.3509\pm 2.2237$ & $2.2117\pm 0.0741$ & $30.0000\pm 0.0000$ \\
ROLeR & \underline{$76.1603\pm 2.1200$} & \underline{$2.5387\pm 0.0707$} & $30.0000\pm 0.0000$ & \underline{$68.3637\pm 1.8550$} & \underline{$2.2788\pm 0.0618$} & $30.0000\pm 0.0000$ \\
\midrule
\rowcolor{lightgray}
DARLR (Ours) & $\mathbf{78.0429 \pm 2.1462}$ & $\mathbf{2.6014 \pm 0.0715}$ & $30.0000 \pm 0.0000$ & $\mathbf{68.5418 \pm 1.9014}$ & $\mathbf{2.2847 \pm 0.0634}$ & $30.0000 \pm 0.0000$ \\
\midrule
\midrule
\textcolor{gray}{GT (Ideal)} & \textcolor{gray}{$80.0895\pm 2.4545$} & \textcolor{gray}{$2.6696\pm 0.0818$} & \textcolor{gray}{$30.0000\pm 0.0000$} & \textcolor{gray}{$68.8791\pm 3.2867$} & \textcolor{gray}{$2.2960\pm 0.1096$} & \textcolor{gray}{$30.0000\pm 0.0000$} \\
\bottomrule
\end{tabular}
}
\end{table*}

\subsubsection{Baselines}
Three categories of baselines are compared in the experiments. For model-based offline RL methods, DeepFM~\cite{deepfm} is employed to learn the default world model. Thus, for each dataset, these methods share the same world model.

\noindent \textbf{Naive methods}:

$\bullet$ \textbf{\textit{$\epsilon$-greedy}} has the greedy policy with maximum predicted reward and a probability of $1-\epsilon$ for random actions.

$\bullet$ \textbf{\textit{UCB}}~\cite{lai1985ucb} estimates confidence intervals and prioritizes actions with higher bounds to balance exploitation and exploration.

\noindent \textbf{Model-free RL}:

$\bullet$ \textbf{\textit{SQN (2020)}}~\cite{xin2020sqn} employs a dual-headed network with cross-entropy loss and RL objectives tailored for RecSys.

$\bullet$ \textbf{\textit{BCQ (2019)}}~\cite{fujimoto2019bcq} focuses on selectively policy update with high-confidence data while excluding ambiguous samples.

$\bullet$ \textbf{\textit{CQL (2020)}}~\cite{kumar2020cql} employs a conservative strategy for out-of-distribution data influence when updating state-action values.

$\bullet$ \textbf{\textit{CRR (2020)}}~\cite{wang2020crr} refines policies by aligning updates with the deviation observed in behavior policies, emphasizing stability.

\noindent \textbf{Model-based RL}:

$\bullet$ \textbf{\textit{IPS (2015)}}~\cite{swaminathan2015ips} applies statistical sample re-weighting to enable actor-critic policies to learn from the re-weighted offline dataset.

$\bullet$ \textbf{\textit{MBPO (2019)}}~\cite{janner2019mbpo} optimizes an actor-critic policy through iterative training on a learned world model.

$\bullet$ \textbf{\textit{MOPO (2020)}}~\cite{yu2020mopo} introduces an uncertainty penalty via an ensemble of world models to improve policy robustness.

$\bullet$ \textbf{\textit{DORL (2023)}}~\cite{gao2023dorl} has an uncertainty penalty with entropy to foster diverse recommendations to deal with the Matthew Effect.

$\bullet$ \textbf{\textit{ROLeR (2024)}}~\cite{yi2024roler} leverages non-parametric soft-label clustering to reshape the reward obtained from the world models.

\noindent \textbf{Ideal situation}:

$\bullet$ \textbf{\textit{GT (Ideal)}} uses the ground-truth reward as an ideal baseline and upper-bound for training recommendation policies with A2C.

\subsubsection{Evaluation and Metric}
To effectively evaluate the performance of DARLR, the following evaluation protocol and metrics are employed to keep consistency with previous work~\cite{gao2023cirs,gao2023dorl,yi2024roler}.

\noindent $\bullet$ $R_\text{tra}$ is the mean cumulative reward in testing episodes. It is a direct reflection of the long-term user satisfactory and user engagement.

\noindent $\bullet$ $R_\text{each}$ represents the mean single-step reward during testing. It shows the overall single-step satisfactory of users.

\noindent $\bullet$ Length is the mean interaction length of the testing trajectories.

\noindent $\bullet$ Majority category domination (MCD) is an indication of the recommendation diversity based on the tags of items proposed by~\cite{gao2023dorl}. Note that (1) MCD is not an absolutely comparable metric for recommendation performance. Thus, when MCD of a method is in a reasonable range, the recommendation diversity is considered reasonable. (2) MCD is only applicable for KuaiRec and KuaiRand but not for Coat and Yahoo due to the availability of tags.

There are two termination conditions: (1) $M$ items of the same category are recommended to a user in its recent $N$ transitions. For a fair comparison with~\cite{gao2023dorl,yi2024roler}, $M=0, N=4$ is retained in the experiments, which means for every four transitions, the items' categories should differ; (2) the maximum interaction length: 30.

\subsubsection{Implementation}
\label{sec:implementation}
The sampling steps for the recommender is $100000$ and the number of reference users of the selector is chosen from $[10, 20, 30, 40]$ for KuaiRec and Coat, $[50, 100, 150, 200]$ for KuaiRand, $[25, 50, 75, 100]$ for Yahoo. Both the similarity gain, \textit{i.e.,} $r^{\text{sel}}_s$, and the diversity gain, \textit{i.e.,} $r^{\text{sel}}_d$ range in $[0, 1]$, and the former is expected to have greater influence on the intrinsic reward. Thus, $\lambda_s$ and $\lambda_d$ are evaluated in $[0.5, 1, 2, 5]$ and $[0.01, 0.05, 0.1, 0.5]$, respectively across all datasets. Then, $\lambda_U$ and $\lambda_E$ are tuned in $[0.01, 0.05, 0.1, 0.5, 1]$ across all datasets. The state transition functions for the selector and recommender are Transformer~\cite{vaswani2017attention} whose number of encoder layers and attention heads are chosen from $[1, 2, 3]$. The window sizes for both transition functions are chosen from $[3, 5, 10]$. Adam is used as the optimizer for the actors and critics of the selector and the recommender, as well as the state transition functions. The default learning rate is set to $0.001$.

Note that while the dual-agent learning paradigm introduces an additional RL loop, the forward sampling efficiency of the selector can be enhanced by constraining the number of reference users and performing user selection within a smaller user subset, typically obtained via clustering techniques. Meanwhile, the backward computation of the dual-agent framework can be executed in parallel, ensuring that the overall computational complexity of DARLR remains comparable to existing approaches like DORL and ROLeR. 
Empirically, all experiments were conducted using an NVIDIA RTX™ A6000 GPU with 48 GB GDDR6 memory, with each trial completed within 10 GPU hours.
\vspace{-0.2em}

\begin{figure}[!t]
\centering
\includegraphics[trim=0cm 0cm 0cm 0cm, clip, width=0.95\columnwidth]{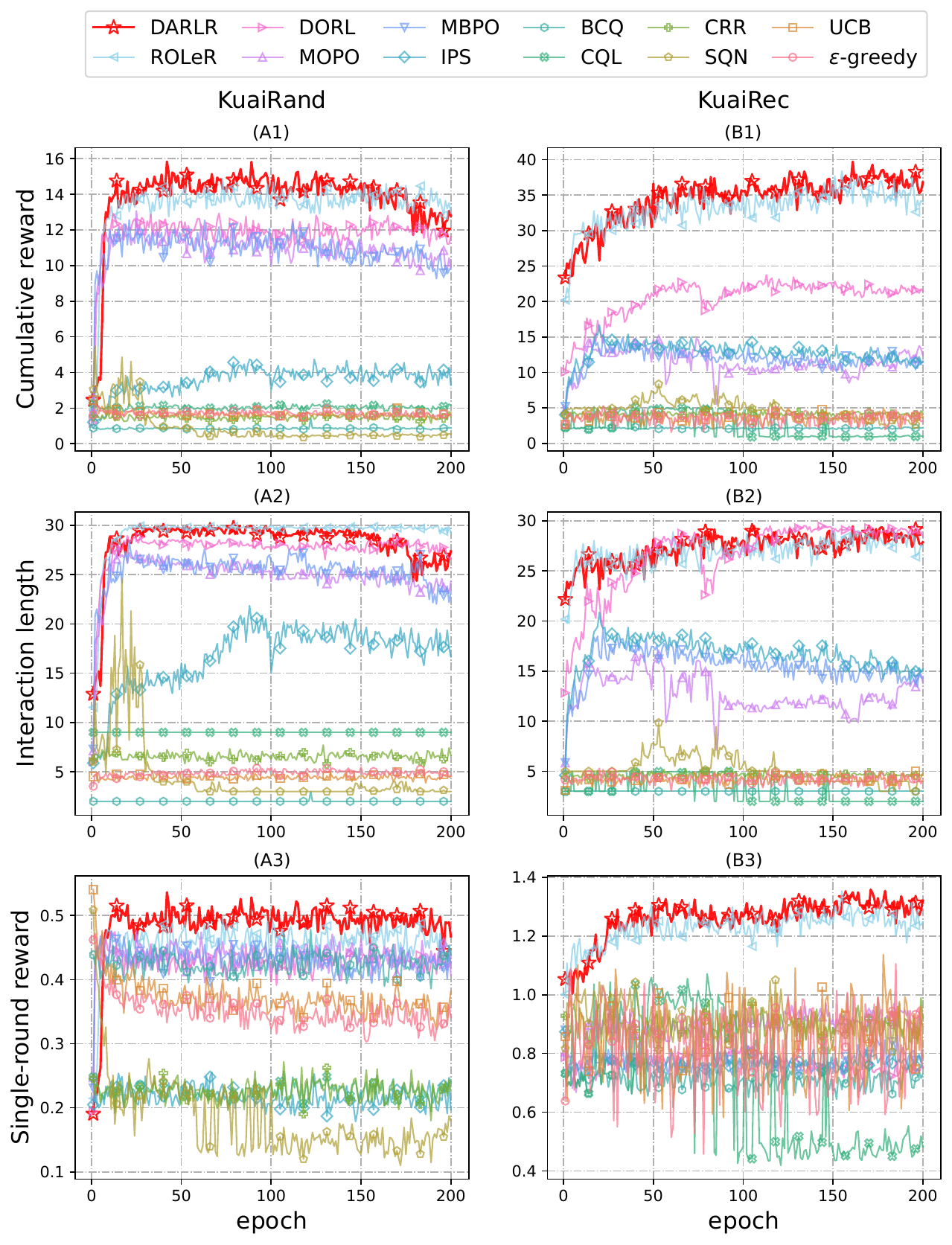}
\vspace{-1em}
\caption{The overall performance on KuaiRand and KuaiRec.}
\label{fig:main_result}
\vspace{-1em}
\end{figure}

\subsection{Overall Performance (RQ1)}
The results of the main experiments on four benchmark datasets are exhibited in Table~\ref{tb:main_result1} and~\ref{tb:main_result2}. For more intuitive observation, the learning curves on KuaiRand and KuaiRec are shown in Figure~\ref{fig:main_result}. From these results, it can be found that the proposed DARLR achieves a better or comparable performance against current state-of-the-art methods with respect to the cumulative reward on all datasets. Meanwhile, DARLR yields the closest performance to the ideal scenarios where the recommendation policies are trained in the testing environments. Considering the termination conditions of the user-RecSys interaction, higher cumulative rewards, \textit{i.e.,} $R_\text{tra}$, comes from a better balance between the exploitation and exploration, which is extra difficult in the offline settings. Purely pursuing a high average single-step reward, \textit{i.e.,} $R_\text{each}$, often leads to premature termination, thereby constraining long-term user engagement. This phenomenon is evident in the performance of MBPO and MOPO on KuaiRand, as well as SQN and CRR on KuaiRec. While these methods achieve relatively high $R_\text{each}$ values, their overall returns are limited by the constrained interaction length \textit{i.e.,} $Length$. Moreover, prioritizing the maximization of $Length$ alone is insufficient to optimize user satisfaction. On Coat and Yahoo, while all methods achieve the maximum interaction length, the single-step reward becomes the dominant factor of the cumulative reward.

Follow the above analyse and break down the cumulative reward into $R_\text{each}$ and $Length$. DARLR achieves the highest single-step reward across the four datasets. Specifically, the primary advantages of both ROLeR and DARLR stem from high $R_\text{each}$. Both methods emphasize the importance of reward model accuracy. While they do not show significant difference on Yahoo as their performance is very close to that of the ideal scenario, DARLR's superior performance on KuaiRec and Coat compared to ROLeR relies on higher single-step rewards. This advantage is due to DARLR's dynamic updating of reward models, enabling more accurate estimations of real-world environments. In addition, while DARLR and ROLeR achieve similar $R_\text{each}$ and interaction length on KuaiRec, DARLR outperforms ROLeR significantly in cumulative reward. This result highlights that DARLR is sophisticated in effectively balancing exploration and exploitation.

Apart from the above analysis, there are some valuable findings. As introduced before, MCD is an associated metric which should be controlled in a certain range instead of being minimized. It can be evident by the MCD of IPS on KuaiRand and KuaiRec. Though IPS has the lowest MCD on the two datasets, the corresponding performance is inferior to strong baselines. By analyzing the MCD and $Length$ of IPS, it is apparent that the relatively low performance on the two datasets is due to an excessive bias toward exploration. Similarly, examining the MCD and $R_{each}$ of BCQ on KuaiRand and SQN on KuaiRec reveals that an overemphasis on exploitation results in early termination, ultimately constraining cumulative rewards. In general, model-based offline RL algorithms such as MOPO, DORL, ROLeR, and DARLR outperform model-free methods like BCQ and CQL, as well as bandit-based approaches, on KuaiRand and KuaiRec. This superiority may be attributed to the higher category density in these datasets, which challenges model-free methods in achieving long interaction lengths. Conversely, on Coat and Yahoo, where reaching interaction limits is less challenging for the baselines, model-free RL algorithms demonstrate competitive performance. However, their effectiveness is still hindered by the inaccuracies and utilization of the reward models.
\vspace{-0.3em}

\subsection{Ablation Study (RQ2)}
The effectiveness of the key designs and the proposed components are verified in this part. Recalling the method of DARLR, the selector enables dynamic reward shaping and uncertainty estimation. Thus, the static one-time reward shaping method, \textit{i.e.,} ROLeR, is used as a baseline. Moreover, DARLR w. $r_{\text{static}}$ and DARLR w. $P_{U,{\text{static}}}$ represent variants that change the reward function and uncertainty penalty to that of the ROLeR, respectively. Looking into the selector, the intrinsic reward design plays a critical role. To test the components of the intrinsic reward, DARLR w. $\hat{r}$ refers to the variant that the selector shares the same reward with the recommender. Further, DARLR w. $\hat{r}+r_s^{sel}$ and DARLR w. $\hat{r}+r_d^{sel}$ denote the variants that the intrinsic rewards are calculated as $\hat{r}+r_s^{sel}$ and $\hat{r}+r_d^{sel}$, respectively. The results are listed in Table~\ref{tab:abla}.

\begin{table}[!t]
\centering
\caption{Ablation study on DARLR's components and design. Evaluation metric: cumulative reward ($R_{\text{tra}}$).}
\vspace{-0.7em}
\label{tab:abla}
\resizebox{1\linewidth}{!}{
\begin{tabular}{c|c|c|c|c}
  \toprule
  Methods & KuaiRec & KuaiRand & Coat & Yahoo \\
  \midrule\midrule
  ROLeR & $33.2457  \pm 2.6403$ & $13.4553 \pm 1.5086 $ & $76.1603 \pm 2.1200$ & $68.3637 \pm 1.8550$ \\
  \midrule
  DARLR $w.\ r_{static}$ & $32.4758 \pm 1.9175$ & $12.9016 \pm 1.6503$ & $74.7500 \pm 2.0629$ & $66.0434 \pm 2.0149$ \\
   DARLR $w.\ P_{U,\text{static}}$ & $32.8984 \pm 1.7779$ & $11.5230 \pm 1.9026$ & $73.7688 \pm 2.0995$ & $66.7788 \pm 3.7383$ \\
   DARLR $w.\ \hat{r}$ & $13.4651 \pm 1.8317$ & $8.6143 \pm 1.6487$ & $69.4249 \pm 1.9675$ & $56.2085 \pm 2.9618$ \\
   DARLR $w.\ \hat{r} + r^{\text{sel}}_s$ & $34.1514 \pm 2.3736$ & $11.7861 \pm 1.4230$ & $75.0405 \pm 2.0078$ & $67.6826 \pm 2.2255$ \\
   DARLR $w.\ \hat{r} + r^{\text{sel}}_d$ & $15.7642 \pm 1.7882$ & $6.7861 \pm 1.4230$ & $67.8568 \pm 2.4669$ & $58.5037 \pm 2.3929$ \\
   \midrule
   DARLR & $35.2203 \pm 2.7576$ & $13.8152 \pm 1.9351$ & $78.0429 \pm 2.1462$ & $68.5418\pm 1.9014$ \\
  \bottomrule
\end{tabular}%
}
\vspace{-1em}
\end{table}

The complete DARLR outperforms its variants and the ROLeR baseline, confirming the advantages of the selector, namely the benefits of its dynamic reward shaping and uncertainty penalties. Looking into the second and third rows in Table~\ref{tab:abla}, both DARLR w. $\hat{r}+r_s^{sel}$ and DARLR w. $\hat{r}+r_d^{sel}$ cannot outperform ROLeR and DARLR, which suggests that the static reward shaping does not cope well with the dynamic uncertainty estimation and vice veasa. While DARLR w. $\hat{r}$ aligns with the selector's intended online reward model (\ref{sec:selector}), its offline performance suffers due to the inability to accurately evaluate reference user selection, inducing biases when sharing the recommender's reward. Similarly, the ineffectiveness of DARLR w. $\hat{r} + r_d^{\text{sel}}$ underscores the limitations of adding diversity-driven intrinsic rewards when reference user evaluation is misaligned. Interestingly, DARLR w. $\hat{r} + r_s^{\text{sel}}$ achieves better outcomes among variants, though still suboptimal compared to ROLeR. This highlights the dual importance of leveraging user similarity for reference selection and incorporating diversity gains in intrinsic reward design, which are critical for robust performance.

\begin{figure}[!t]
\centering
\begin{minipage}[b]{0.48\columnwidth}
    \includegraphics[trim=0cm 0cm 0cm 0cm, clip, width=\columnwidth]{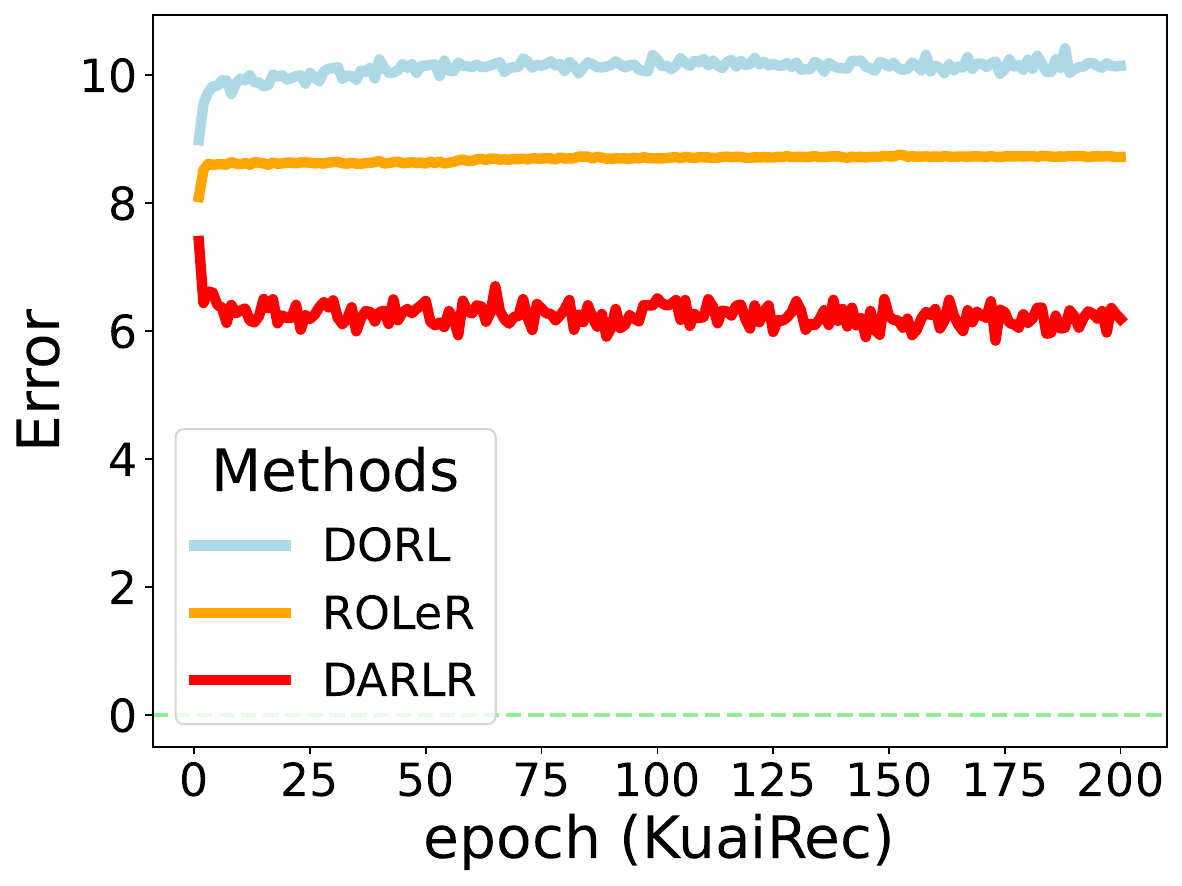}
    \vspace{-2.5em}
\end{minipage}
\hfill
\begin{minipage}[b]{0.48\columnwidth}
    \includegraphics[trim=0cm 0cm 0cm 0cm, clip, width=\columnwidth]{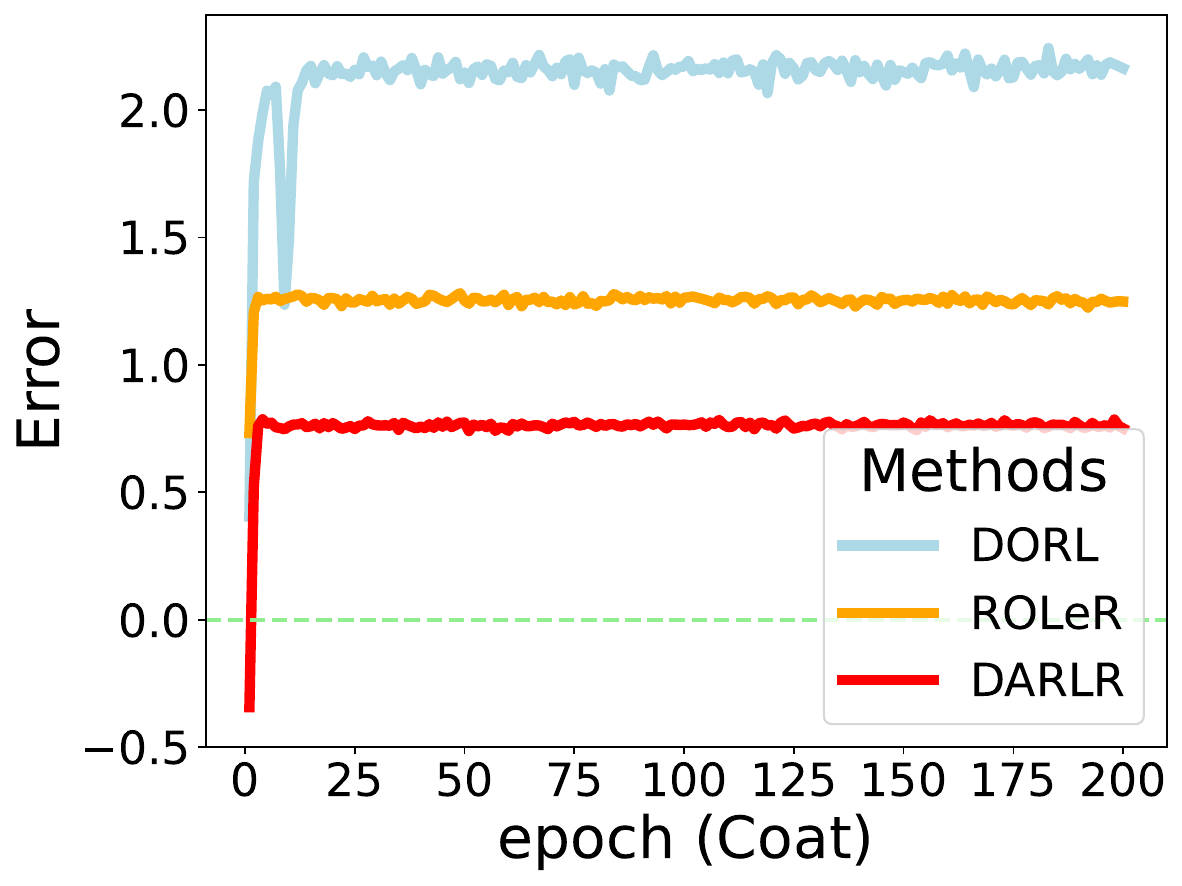}
    \vspace{-2.5em}
\end{minipage}
\caption{The mean reward differences (Error) during training of DORL, ROLeR and DARLR. The reward differences of the dynamic reward shaping method (DARLR) are consistently smaller than those of the static reward shaping method (ROLeR and DORL).} 
\label{fig:rq3}
\vspace{-1em}
\end{figure}

\subsection{Effectiveness of Dynamic Reward (RQ3)}
This section analyses the impact of dynamic reward shaping during policy learning. The mean reward differences per training epoch (“Error”), defined as the deviation between ground truth and estimated rewards, are evaluated on KuaiRec and Coat datasets. DORL and ROLeR, with static reward shaping, serve as baselines. 

As shown in Figure~\ref{fig:rq3}, DARLR, leveraging dynamic reward shaping, consistently achieves lower reward differences compared to DORL and ROLeR across all training epochs in both environments. These findings align with the recommendation performance reported in Tables~\ref{tb:main_result1} and~\ref{tb:main_result2}, further highlighting the efficacy of dynamically updating the reward model. In contrast, the relatively flat reward difference levels of the other methods underscore the limitations of static reward shaping, which struggles to capture the evolving nature of recommendation tasks. Notably, the reward difference curves on Coat (right plot in Figure~\ref{fig:rq3}) exhibit sudden increases during the initial epochs, potentially caused by sampling on accurately estimated interactions as all three curves share the same phenomenon with the same seed. Despite this, DARLR maintains competitive reward differences, ranging between $(0.5, 1)$, demonstrating its robustness across all epochs.

\subsection{Hyperparameter Sensitivity (RQ4)}
\begin{figure}[!t]
\centering
\begin{minipage}[b]{0.48\columnwidth}
    \includegraphics[trim=0cm 0cm 0cm 0cm, clip, width=\columnwidth, height=3.25cm]{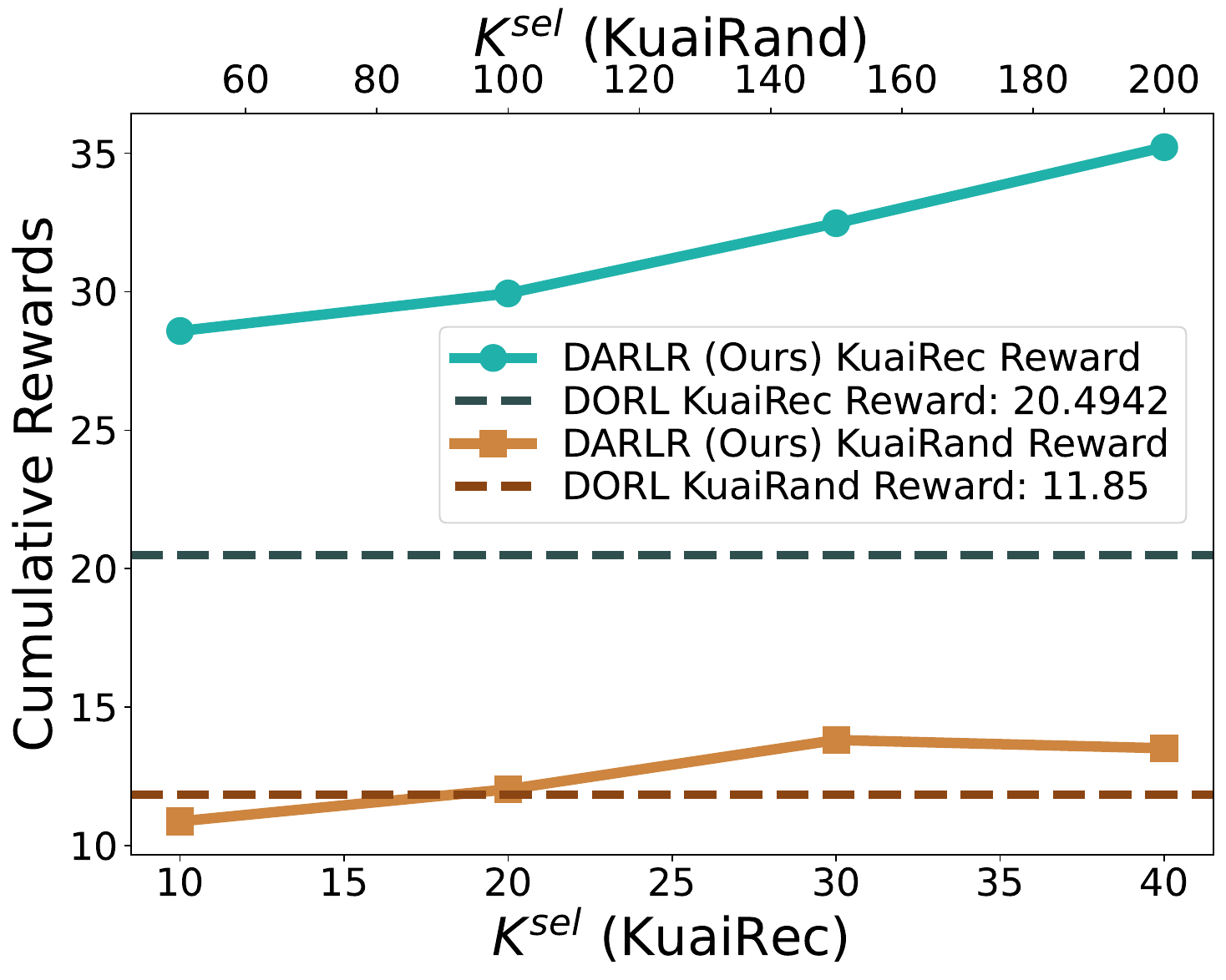}
    \vspace{-2.75em}
\end{minipage}
\hfill
\begin{minipage}[b]{0.48\columnwidth}
    \includegraphics[trim=0cm 0cm 0cm 0cm, clip, width=\columnwidth, height=3.25cm]{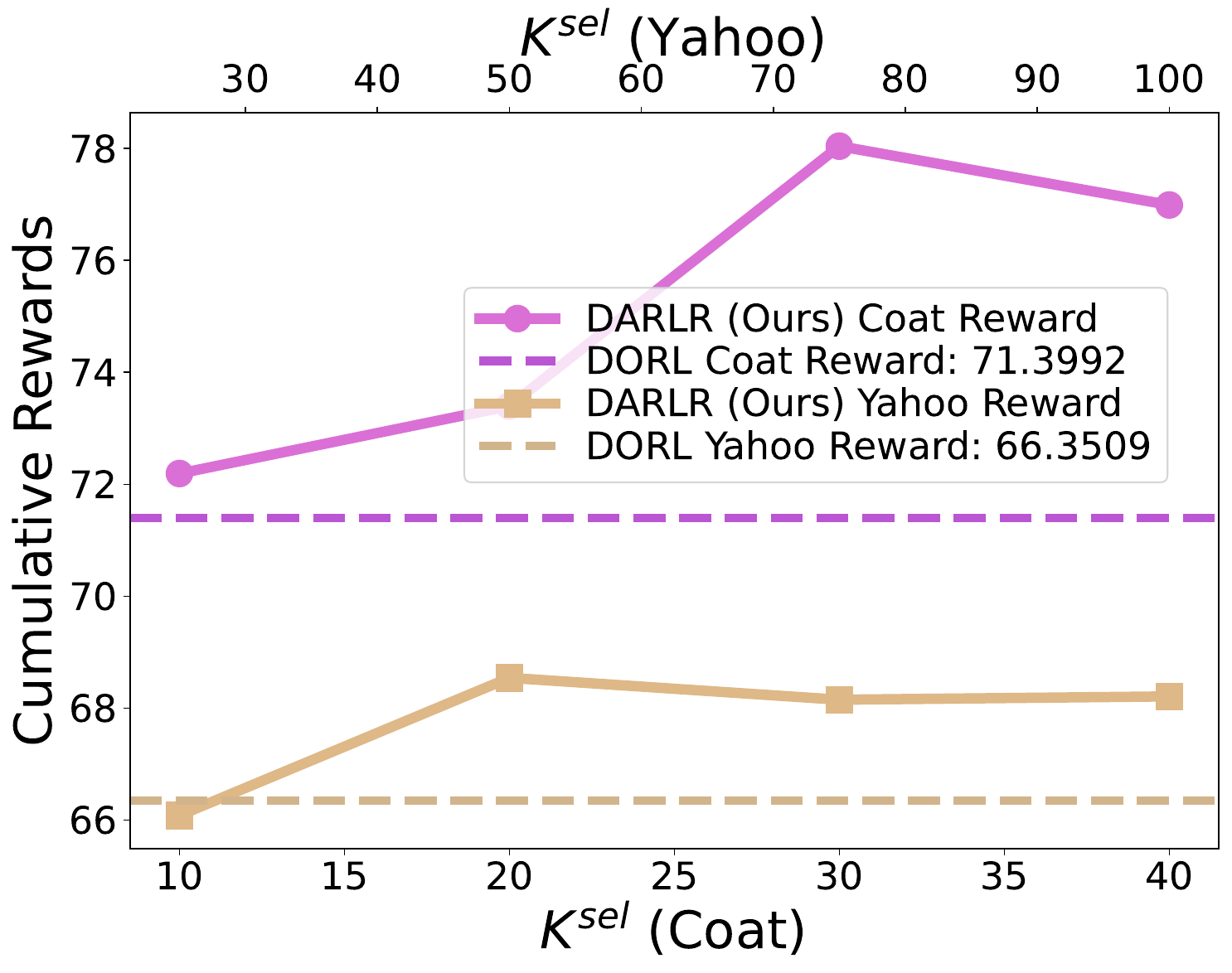}
    \vspace{-2.75em}
\end{minipage}
\vspace{1.5em}

\begin{minipage}[b]{0.48\columnwidth}
    \includegraphics[trim=0cm 0cm 0cm 0cm, clip, width=\columnwidth, height=3.25cm]{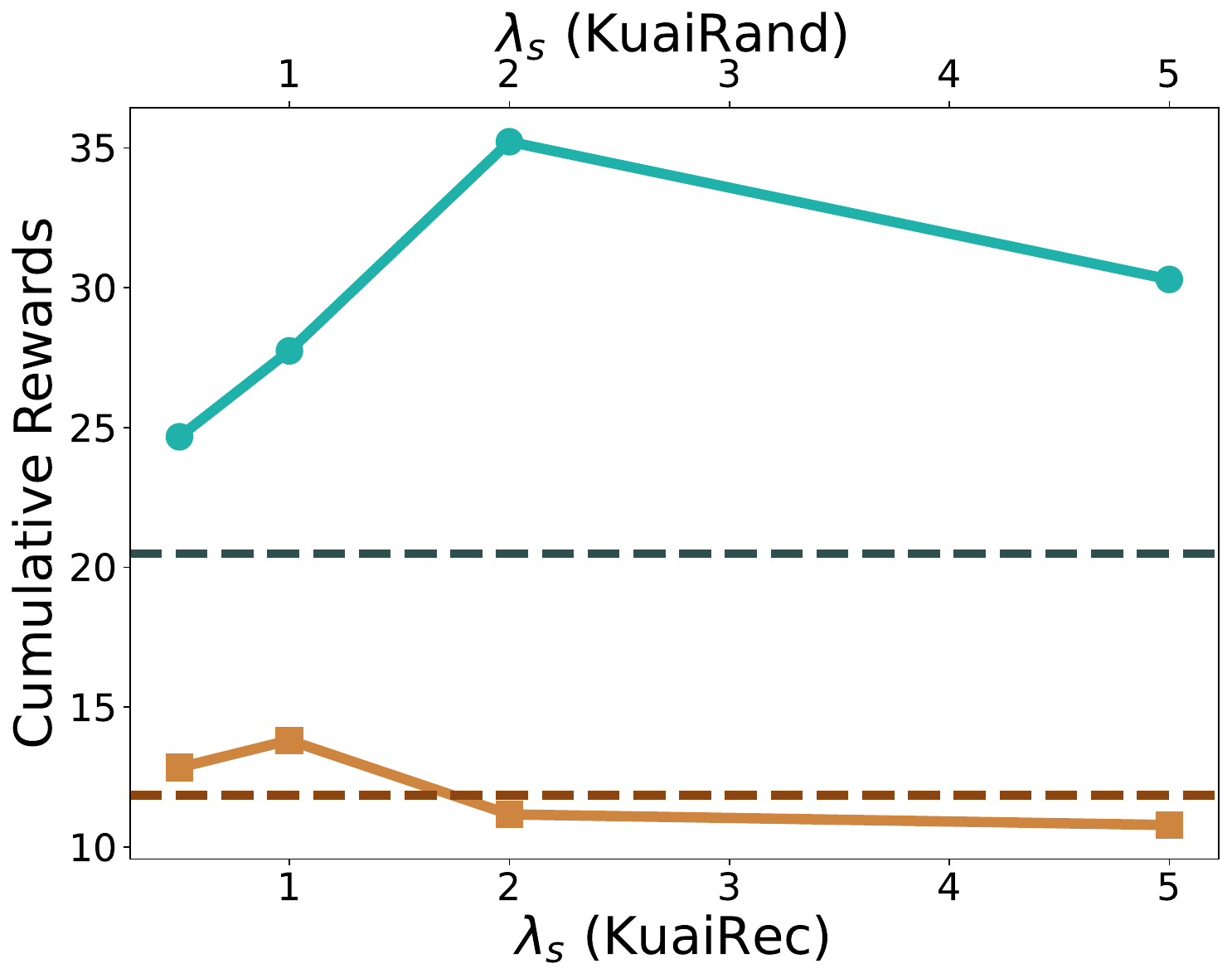}
    \vspace{-2.75em}
\end{minipage}
\hfill
\begin{minipage}[b]{0.48\columnwidth}
    \includegraphics[trim=0cm 0cm 0cm 0cm, clip, width=\columnwidth, height=3.25cm]{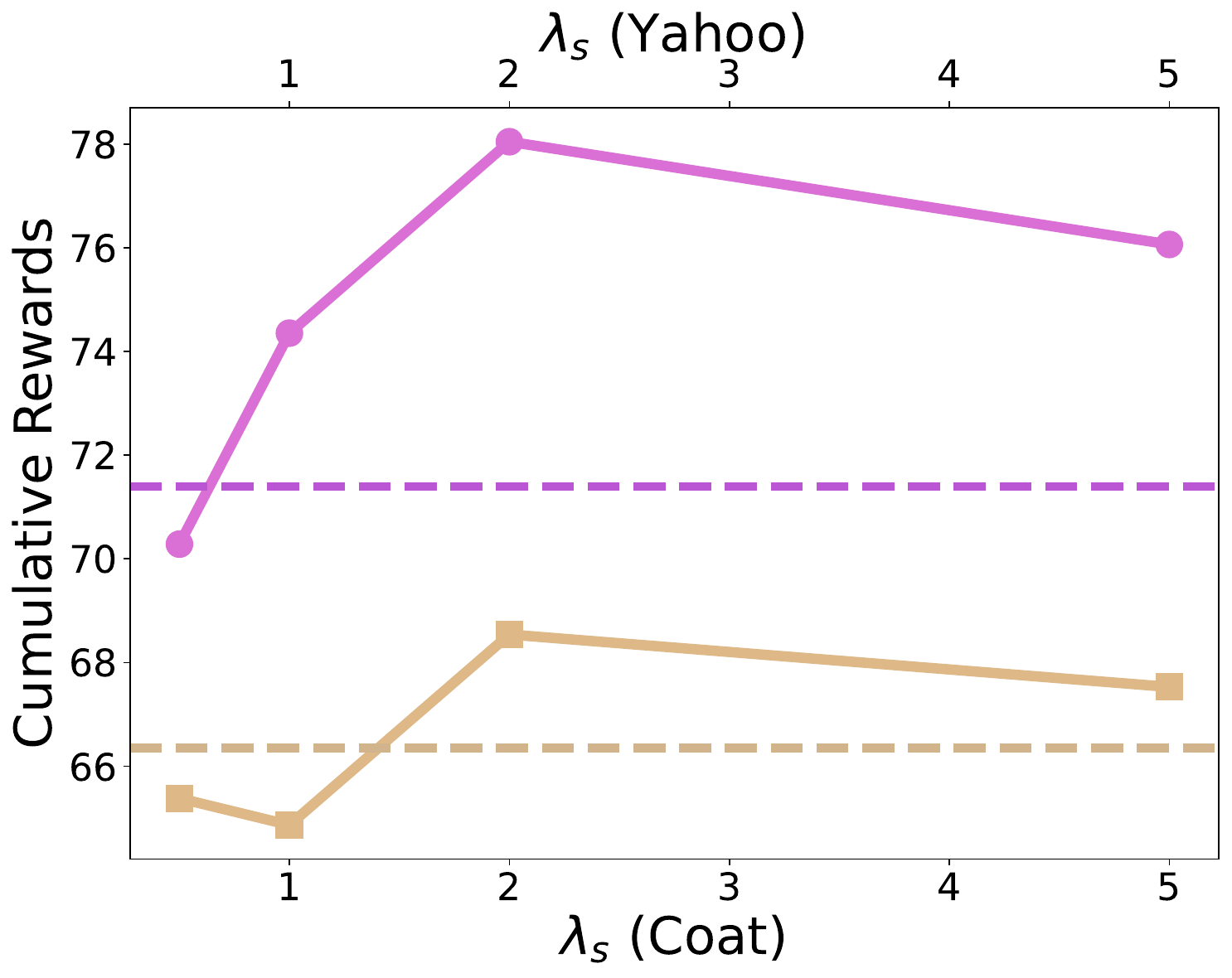}
    \vspace{-2.75em}
\end{minipage}
\vspace{1.5em}

\begin{minipage}[b]{0.48\columnwidth}
    \includegraphics[trim=0cm 0cm 0cm 0cm, clip, width=\columnwidth, height=3.25cm]{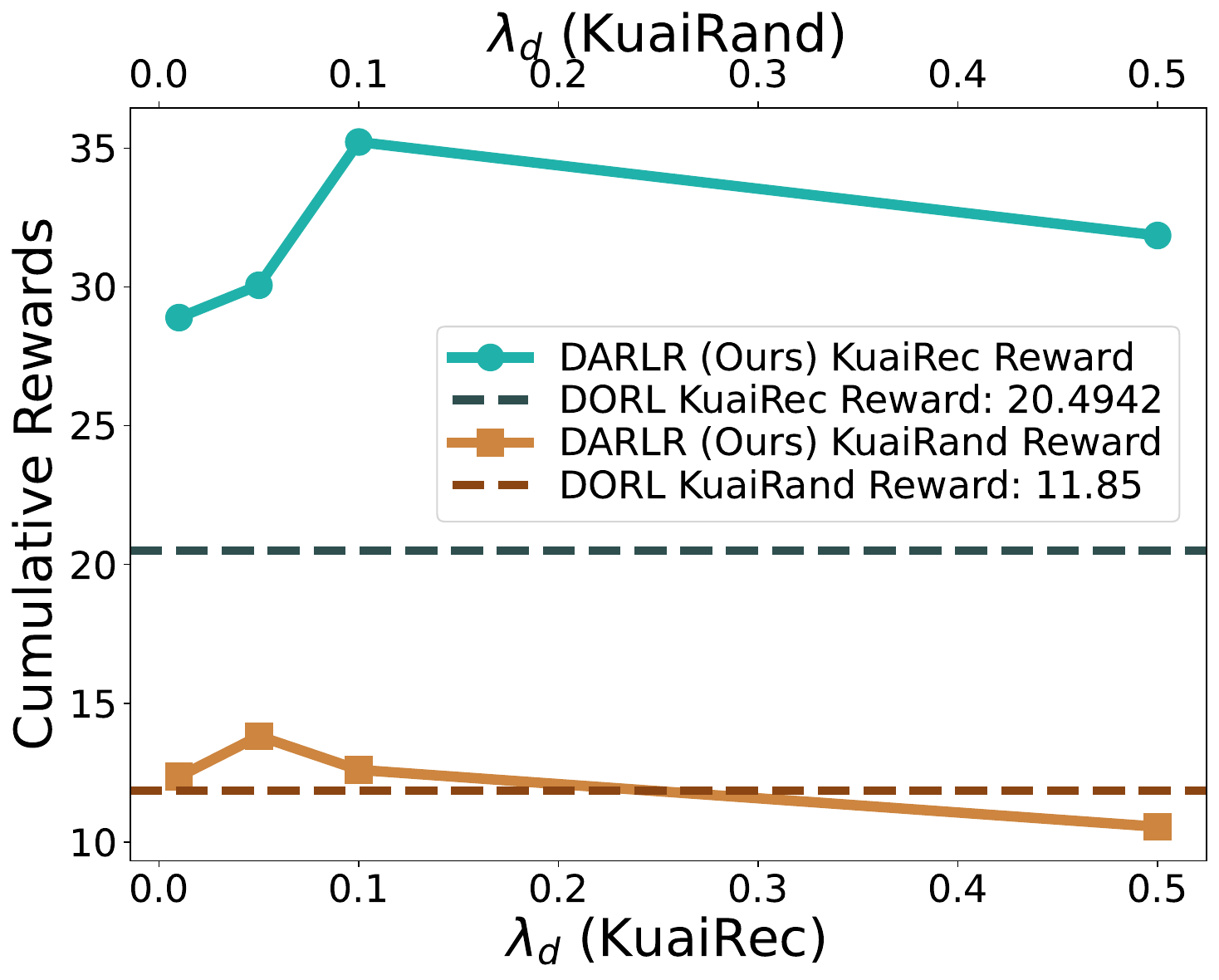}
    \vspace{-2.75em}
\end{minipage}
\hfill
\begin{minipage}[b]{0.48\columnwidth}
    \includegraphics[trim=0cm 0cm 0cm 0cm, clip, width=\columnwidth, height=3.25cm]{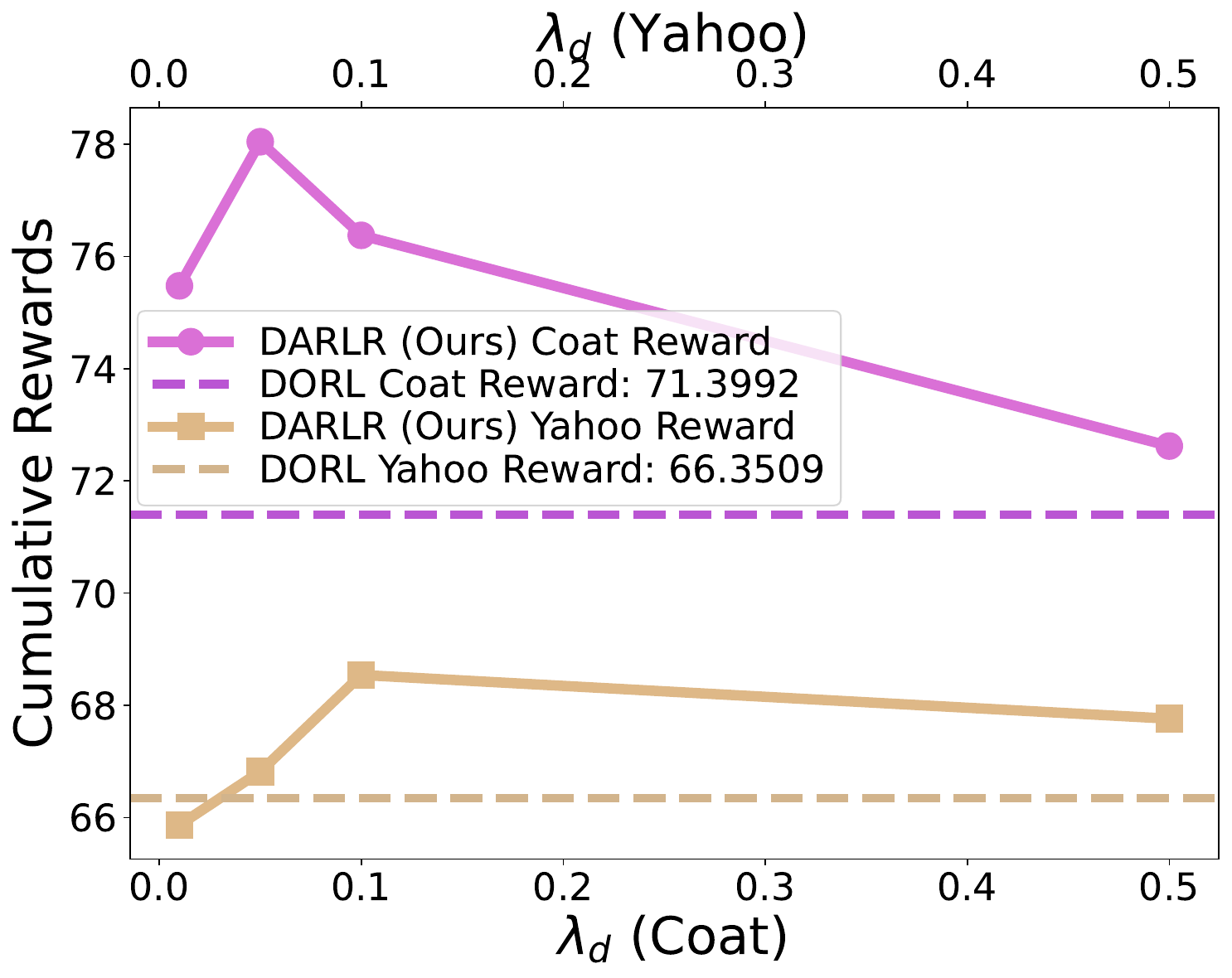}
    \vspace{-2.75em}
\end{minipage}
\vspace{1.5em}

\begin{minipage}[b]{0.48\columnwidth}
    \includegraphics[trim=0cm 0cm 0cm 0cm, clip, width=\columnwidth, height=3.25cm]{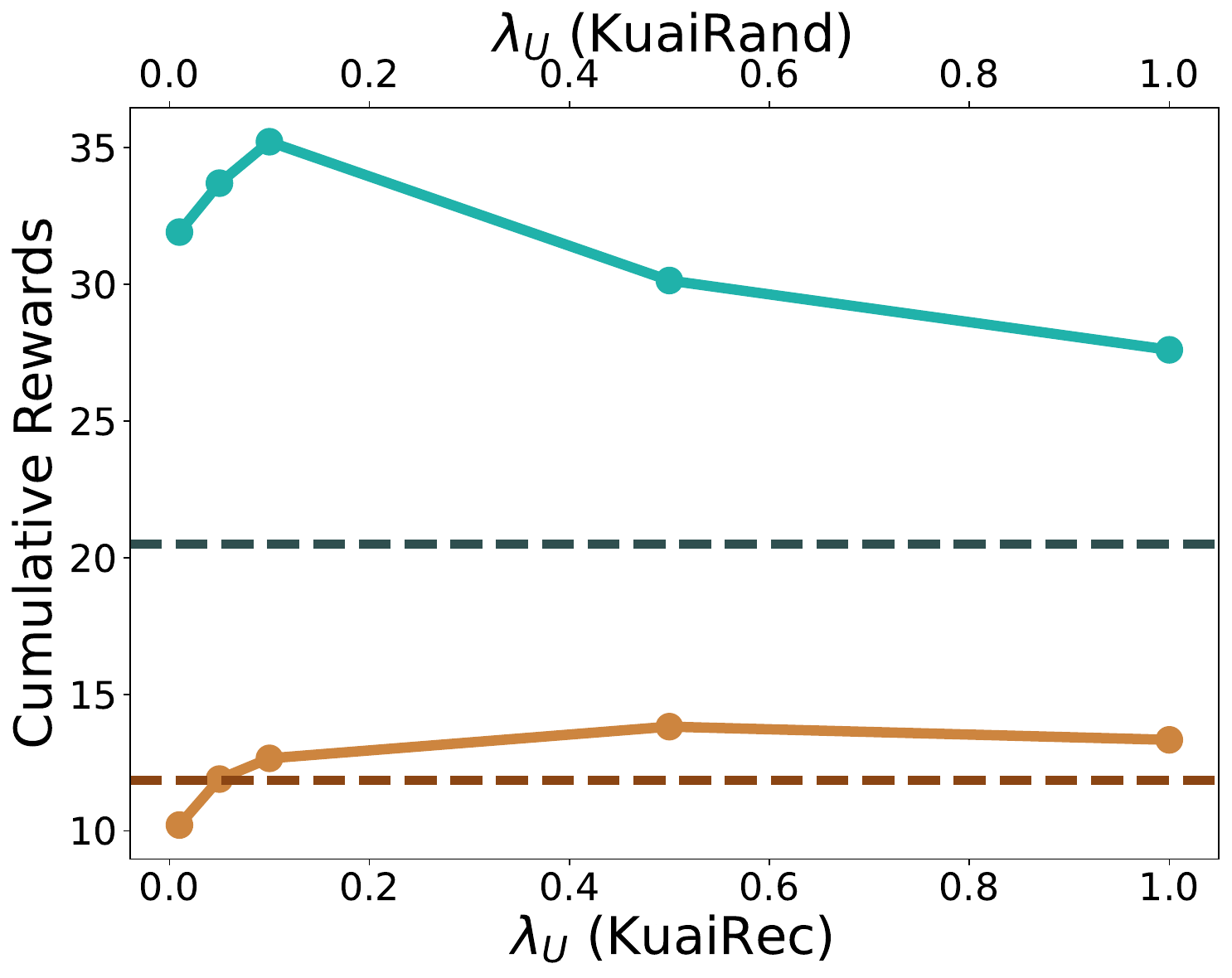}
    \vspace{-2.75em}
\end{minipage}
\hfill
\begin{minipage}[b]{0.48\columnwidth}
    \includegraphics[trim=0cm 0cm 0cm 0cm, clip, width=\columnwidth, height=3.25cm]{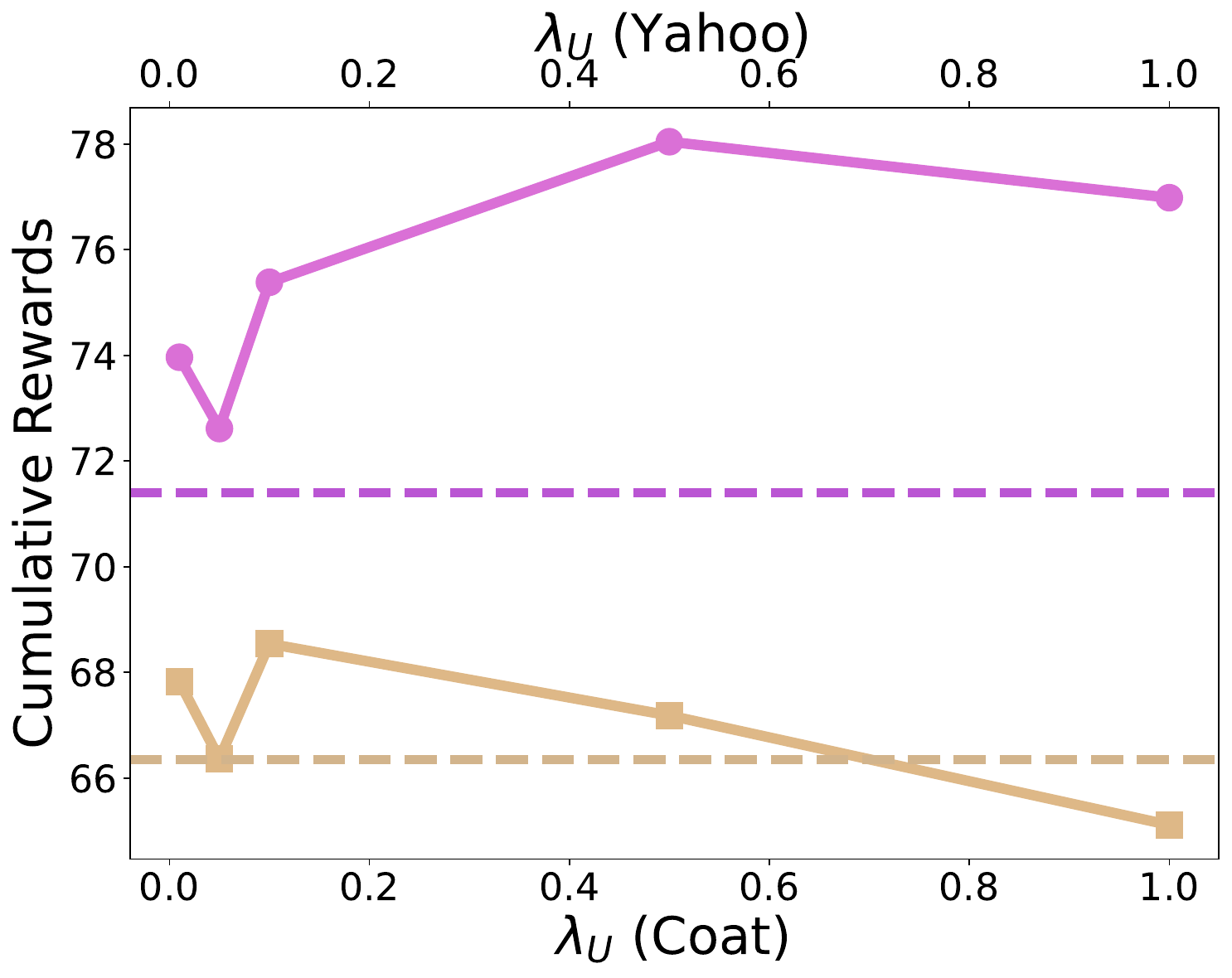}
    \vspace{-2.75em}
    \end{minipage}
\vspace{1.5em}

\begin{minipage}[b]{0.48\columnwidth}
    \includegraphics[trim=0cm 0cm 0cm 0cm, clip, width=\columnwidth, height=3.25cm]{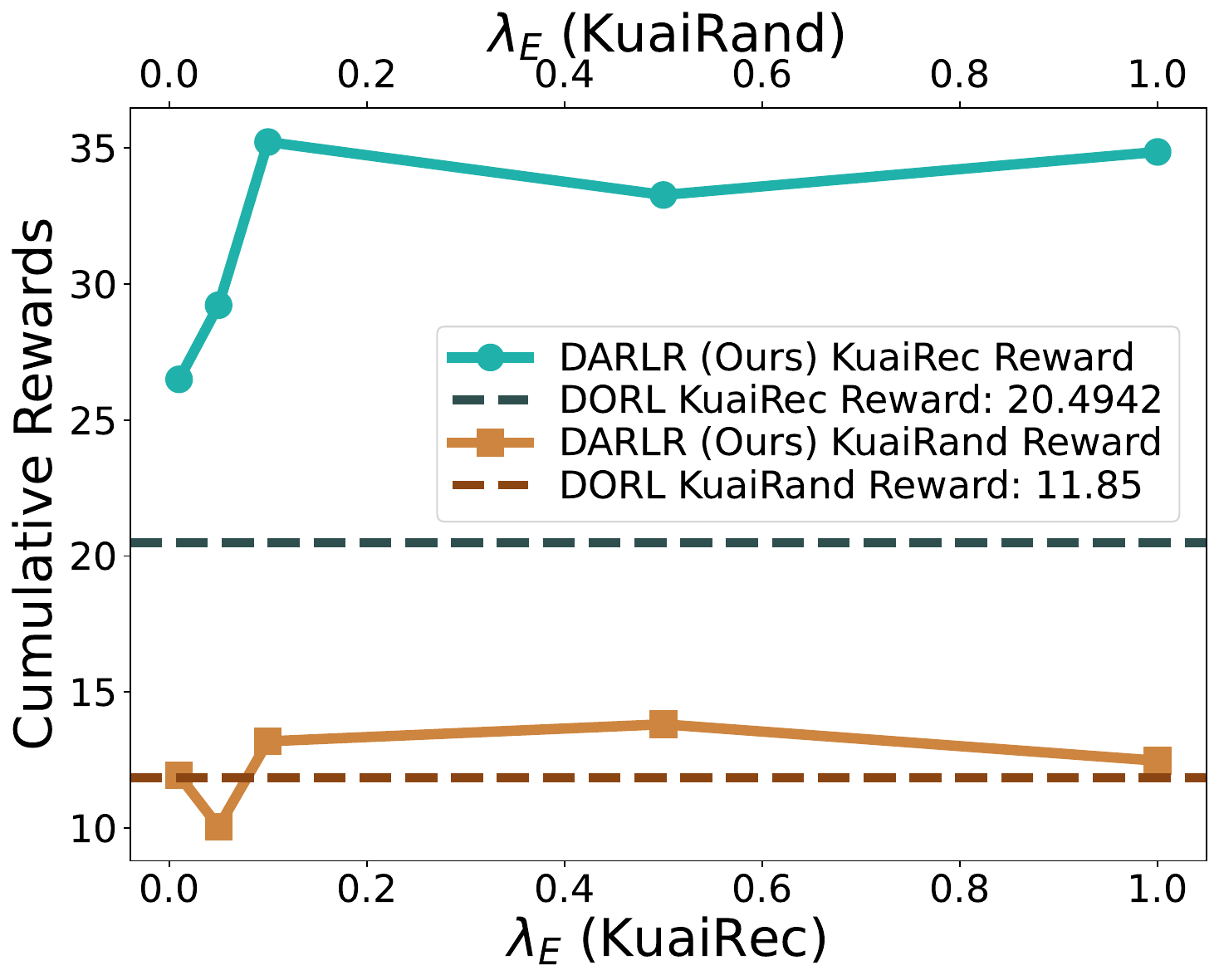}
    \vspace{-2.3em}
\end{minipage}\hfill
\begin{minipage}[b]{0.48\columnwidth}
    \hspace{-0.3em}
    \raisebox{0pt}{\hspace{-0.3em}%
    \includegraphics[width=\linewidth, trim=0cm 0cm 0cm 0cm, clip]{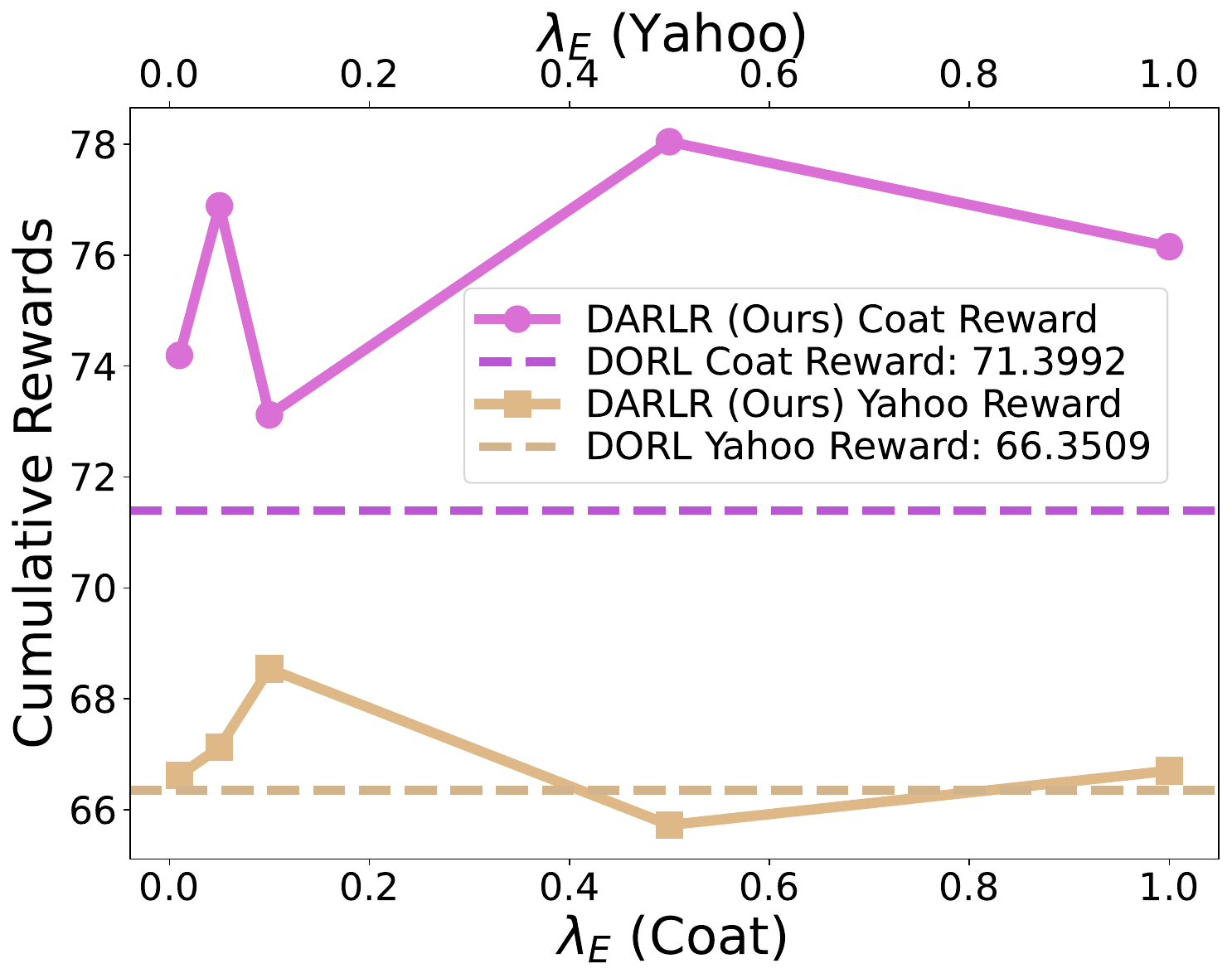}}
    \vspace{-1em}
\end{minipage}
\caption{Hyperparameter sensitivity with different $K^{\text{sel}}$, $\lambda_s$, $\lambda_d$, $\lambda_{U}$, and $\lambda_{E}$ on four datasets.}
\label{fig:robust}
\vspace{-1.5em}
\end{figure}

For DARLR, there are five key hyperparameters: (1) $K^{\text{sel}}$, the termination of the selection process described in Sec.\ref{sec:selector}; (2) $\lambda_s$, the coefficient of the similarity gain in the intrinsic reward in Eq. (\ref{eq:intrinsic_r}) and (3) $\lambda_d$, the coefficient of the diversity gain; (4) $\lambda_U$, the coefficient of the uncertainty penalty in Eq. (\ref{eq:rec_reward}) and (5) $\lambda_E$, the coefficient of the entropy penalty. The testing ranges of hyperparameters have been described in Sec.~\ref{sec:implementation}. The corresponding results are listed in Fig.\ref{fig:robust}. The cumulative reward is used as the evaluation metric and the dash line in each subplot represents the performance of DORL. Since the subplots in each column share the same legends, part of them are omitted for brevity. 

Observing the first row in Fig.\ref{fig:robust}, $K^{\text{sel}}$ significantly influences performance, with dataset-specific optimal values. Smaller $K^{\text{sel}}$ benefits denser datasets such as KuaiRec, while larger $K^{\text{sel}}$ enhances performance on sparser datasets like KuaiRand. The second row shows the influence of the similarity gain's coefficient ($\lambda_s$). Obvious peaks can be found in the subplots, similar to the observation in the third row--the subplots for the coefficient of the diversity gain ($\lambda_d$). Combing the results of the four subplots, both $\lambda_s$ and $\lambda_d$ have a significant impact on the cumulative reward, and the scale of $\lambda_s$ is supposed to be larger. In addition, according to the last two rows in this figure, both the coeffients of the uncertainty penalty and entropy penalty yield the highest cumulative reward at moderate values, emphasizing the importance of the balance of reward maximization, uncertainty handling and exploration robustness.

\section{Related Work}
\noindent $\bullet$ \textbf{Reinforcement Learning in Recommender System.}
Given the interactive nature and sparsity of RecSys~\cite{qiu2022contrastive,qiu2021memory}, RL methods draw increasing attention~\cite{zhang2019deep,chen2023deep}. With the MDP formalization, some researches explore to derive informative state representations from user features and historical interactions~\cite{kang2018sasrec,huang2022state_repr}. SlateQ~\cite{ie2019slateq} proposes to decompose the large action space in Q-learning based methods. PrefRec~\cite{wanqi2023prefrec} utilizes RLHF~\cite{stiennon2020rlhf} to learn a reward model from sparse datasets. For the practicability,~\cite{chen2019top} adapts REINFORCE~\cite{williams1992reinforce} to a large action space on the orders of millions. ~\cite{chen2021user} augments the policy learning to improve sample efficiency, targeting the sparse signal issue in RecSys. 
Some methods explore offline RL in RecSys. CIRS~\cite{gao2023cirs} uses causal graphs to capture the user preferences. DORL~\cite{gao2023dorl} introduces entropy penalty to encourge the exploration of offline policies. ROLeR~\cite{yi2024roler} proposes a static reward shaping method to ease the impact of inaccurate world models.
DARLR operates within the same offline model-based RL setting, yet the significant difference lies in the evolving world model in contrast to the fixed models applied in these related works.

\noindent $\bullet$ \textbf{Multi-agent Reinforcement Learning in Recommender System.}
A multi-agent system (MAS) contains more than one intelligent agents~\cite{busoniu2008marl_survey} with wide applications in robotic control~\cite{peng2021facmac}, traffic flow management~\cite{zhang2019cityflow}, and video games~\cite{lowe2017maddpg,rashid2020qmix}. In some scenarios where one agent is incapable of handling the user interactions~\cite{zhang2017dynamic}, user recommendation customization~\cite{zhao2020deepchain}, large hierarchical action spaces~\cite{chen2019large}, etc., MARL exhibits its potentials~\cite{chen2023deep}. MA-RDPG~\cite{feng2018learning} uses two agents' communication to facilitate multi-scenario recommendations. MAHRL~\cite{zhao2020mahrl} uses hierarchical RL to decompose tasks into sub-tasks. RAM~\cite{zhao2020jointly} designs two agents for advertising and recommendation. Nevertheless, the effectiveness of MARL remains underexplored in model-based offline RL for RecSys. 

\noindent $\bullet$ \textbf{Offline Reinforcement Learning.}
Offline RL can train a policy purely from offline data~\cite{levine2020offline,prudencio2023survey,agarwal2020optimistic,kumar2019stabilizing,wang2020statistical}. This learning paradigm is promising in domains where the offline data is significantly less expensive than online interactions such as robotics~\cite{sinha2022s4rl,luo2023action}, autonomous driving~\cite{fang2022offline,diehl2023uncertainty}, and recommender systems~\cite{chen2023opportunities,afsar2022rsrl}. 
Existing model-free offline RL algorithms~\cite{fujimoto2019bcq,kumar2020cql,wang2020crr,ran2023policy} introduces conservatism during training to deal with value overestimation issues, such as BCQ~\cite{fujimoto2019bcq}, CQL~\cite{kumar2020cql}, and PRDC~\cite{ran2023policy}.
Offline model-based RL algorithms~\cite{janner2019mbpo,yu2020mopo,rigter2022rambo,kidambi2020morel,liu2024micro} are more sample-efficient due to the explicit modeling of transition functions and reward functions. MOPO~\cite{yu2020mopo} and MOReL~\cite{kidambi2020morel} learn an ensemble of world models to estimate the state-action uncertainty. RAMBO~\cite{rigter2022rambo} trains an adversarial environment model for RL learning. 
While these methods demonstrate improved performance on RL benchmarks~\cite{justin2020d4rl}, their effective adaptation to RecSys requires further investigation~\cite{chen2023opportunities}.
\vspace{-0.8em}

\section{Conclusion and Future Work}
\vspace{-0.2em}
In this paper, the limitations of frozen reward shaping and uncertainty estimation derived from world models in model-based offline recommender systems are identified. To address these issues, a dual-agent method, DARLR, is proposed to dynamically improve the reward function and uncertainty estimation. Specifically, a selector is employed to select reference users which considers both the similarity and diversity among users. Then, a recommender aggregates the information of reference users to improve the reward functions as well as estimating the uncertainty penalty, facilitating effective offline recommendation policy learning. Empirically,The proposed method is validated through extensive experiments on four challenging datasets, outperforming all baselines, including state-of-the-arts. In future work, the development of dynamic reward shaping methods tailored for large-scale datasets will be investigated. Further, to achieve more efficient hyperparameter tuning and reduce manual effort, the adoption of large language models (LLMs)~\cite{zhang2023llm_hyper} and meta-learning~\cite{vettoruzzo2024meta_learning_survey} in DARLR will be explored.
\vspace{-0.8em}

\section{Acknowledegments}
\vspace{-0.2em}
This work is supported by projects DE200101610, DE250100919, CE200100025 funded by Australian Research Council, and CSIRO’s Science Leader Project R-91559.

\bibliographystyle{ACM-Reference-Format}
\bibliography{sample-base}

\end{document}